\pdfoutput=1
\documentclass[preprint,aps,prd,nofootinbib]{revtex4-2}
\usepackage{amssymb,amsmath,graphicx,color,microtype,enumerate,slashed,hyperref,xcolor,cases,verbatim}
\usepackage{tikz}
\hypersetup{linktocpage=true,colorlinks=true,linkcolor=blue,filecolor=blue,urlcolor=blue, citecolor=blue}

\begin{document}
\title{Decoherence of Cosmological Perturbations from Boundary Terms and the Non-Classicality of Gravity}
\author{Chon Man Sou$^{1,2}$}
\email{cmsou@connect.ust.hk}
\author{Duc Huy Tran$^{1}$}
\email{dhtranaa@connect.ust.hk}
\author{Yi Wang$^{1,2}$}
\email{phyw@ust.hk}
\affiliation{${}^1$Department of Physics, The Hong Kong University of Science and Technology, \\
	Clear Water Bay, Kowloon, Hong Kong, P.R.China}
\affiliation{${}^2$The HKUST Jockey Club Institute for Advanced Study, The Hong Kong University of Science and Technology, \\
	Clear Water Bay, Kowloon, Hong Kong, P.R.China}
\begin{abstract}

We note that the decoherence of inflationary curvature perturbation $\zeta$ is dominated by a boundary term of the gravity action. Although this boundary term cannot affect cosmological correlators $\left\langle \zeta^n \right\rangle$, it induces much faster decoherence for $\zeta$ than that of previous calculations. The gravitational origin of inflationary decoherence sheds light on the quantum (or non-classical) nature of gravity. By comparing with a Schr\"odinger-Newton toy model of classical gravity, we show that gravity theories of classical or quantum origins can be distinguished by comparing their different impacts on decoherence rate of $\zeta$. Our calculation also indicates that density fluctuation $\delta\rho$ better preserves quantum information than $\zeta$ for the purpose of constructing cosmological Bell-like experiments.

\end{abstract}
\maketitle
\tableofcontents

\section{Introduction}
As the leading paradigm of describing the early universe, inflation proposes that the large scale structures in today's universe originate from the enlargement of vacuum fluctuations during a rapidly expanding period \cite{Guth:1980zm,Linde:1981mu,Starobinsky:1982ee,Albrecht:1982wi}. With such a quantum origin, it is expected that cosmological perturbations can be described by a quantum state and experience a process transiting them to today's stochastic distribution. 

The inflationary quantum state has been studied with various concepts from quantum information \cite{Martin:2019wta}, including the squeezed-state description \cite{Grishchuk:1990bj,Albrecht:1992kf}, the Bell inequality \cite{Campo:2005qn,Choudhury:2016cso,Martin:2017zxs}, the quantum discord \cite{Lim:2014uea,Martin:2015qta,Kanno:2016gas}, quantum measurement problem \cite{Martin:2012pea} and the entanglement entropy \cite{Brahma:2020zpk}, as well as the probe with the quantum non-Gaussian correlators \cite{Green:2020whw}. Such details of the quantumness also motivate us to consider why they fit the classical description of cosmological perturbations, suggesting that there is a quantum-to-classical transition. The transition is often manifested with the environment-induced decoherence studied with various tools, including the mean field \cite{Calzetta:1995ys}, the Schr\"odinger wave functional \cite{Nelson:2016kjm}, the master equation \cite{Hu:1992ig,Lombardo:2005iz,Kiefer:2006je,Kiefer:2008ku,Burgess:2006jn,Burgess:2014eoa,Martin:2018zbe}, the related EFT \cite{Boyanovsky:2015tba,Burgess:2022nwu} and the entropy increased by interactions \cite{Friedrich:2019hev}, whereas there are also some early works arguing that the transition can happen without any environment \cite{Guth:1985ya,Polarski:1995jg,Lesgourgues:1996jc,Kiefer:1998qe} as well as the approaches with the non-equilibrium entropy \cite{Brandenberger:1992jh,Brandenberger:1992sr,Prokopec:1992ia}. Focusing on the minimal environment-induced decoherence for the cosmological perturbations, the process is accompanied by the entanglement between the environment and the system, represented by partitioning the scalar curvature modes $\zeta_{\bf k}$ based on their comoving momenta, and their couplings are dominated by gravitational interactions (at least by counting the slow-roll orders of the interactions).

The entanglement caused by the gravitational interaction implies a non-trivial fact that gravity, as a mediator between the system and environment, cannot be classical according to the Local Operations and Classical Communication (LOCC) \cite{Horodecki:2009zz} (see also \cite{Galley:2020qsf} for a proof with the Generalized Probabilistic Theory). Utilizing this fact, there have been some proposals to probe the non-classicality of gravity in laboratory through the entanglement \cite{Bose:2017nin,Marletto:2017kzi,Marshman:2019sne,Bose:2022uxe}. However, the LOCC and the argument of quantum mediator may not be directly applicable to the cosmological perturbations since they are not spatially separated, and even a spectator field in the classical inflationary background evolves as a two-mode squeezed states, entangling the two particles with opposite momenta \cite{Martin:2019wta}. The non-Gaussianity of quantum states is one of the alternative laboratory tests not involving the quantum mediator, since the quantum gravity is expected to induce operators beyond the second order, whereas the semi-classical gravity remains quadratic \cite{Howl:2020isj}. For the cosmological perturbations, the non-classical gravity is manifested by two aspects: \begin{enumerate}
\item Whether the comoving curvature perturbation $\zeta$ can describe the quantum degree of freedom.
\item Whether the ADM constraints can be considered as quantum operator equations.
\end{enumerate}
For the non-classical gravity, the two aspects are clearly {\it yes}, whereas the semi-classical gravity does not support such properties since the metric fluctuations are not quantized, implying that the quantum fluctuation has to be inflaton $\varphi$, and its non-Gaussianity is attributed to the self-interaction \footnote{For the semi-classical gravity, the ADM constraints are equal to the expectation values of linear operators which are vanishing, so the inflaton $\varphi$ acts as a spectator field.}. Decoherence can reflect the non-Gaussianity of quantum states, providing us a quantitative way to compare the quantum states affected by the non-classical gravity and by the self-interaction of inflaton only. 

Previous works of studying the coupling between system and environment with the gravitational nonlinearities consider only the bulk's Lagrangian, and the leading cubic term is \cite{Nelson:2016kjm,Brahma:2020zpk}
\begin{align}
\mathcal{L}_{\rm int}&=-\frac{M_p^2}{2}\epsilon(\epsilon+\eta)a(t)\zeta^2\partial^2\zeta \ , \label{eq:L_int_cubic_usual}
\end{align} 
whereas the other cubic terms involving $\dot{\zeta}\propto\Pi/a^3$ are highly suppressed in terms of the conjugated momentum $\Pi$. Besides this term, it is well-known that there is a boundary term in the Lagrangian, obtained from the ADM formalism \cite{Arnowitt:1962hi}, dominates for super-horizon modes \cite{Maldacena:2002vr,Arroja:2011yj}
\begin{align}
\mathcal{L}_{\rm bd}&=\partial_t\left(-2a^3H M_p^2 e^{3\zeta}\right) \ ,
\end{align}
and it does not contribute to any correlator $\langle \zeta^n 
\rangle$ \footnote{This boundary term can affect correlators involving $\dot\zeta$, such as $\left\langle \dot\zeta^3 \right\rangle$, see also the appendix B of \cite{Celoria:2021cxq} discussing the relationship of conjugate momentum and total-derivative terms in the Lagrangian. But these correlators involving $\dot\zeta$ quickly vanishes on super-Hubble scales.}, so it is often neglected in literature \cite{Arroja:2011yj,Burrage:2011hd}. We will show that such a boundary term without any slow-roll suppression can contribute a non-Gaussian phase to the Schr\"odinger wave functional and non-vanishing correction to the reduced density matrix after tracing out the environment, leading to much larger decoherence rate, compared to the one obtained with (\ref{eq:L_int_cubic_usual}). This result is the main improvement to the calculation in \cite{Nelson:2016kjm}.

The paper is organized as follows. In Sect.~\ref{sec:wave functional}, we first review the formalism in \cite{Nelson:2016kjm} which calculates the decoherence rate of cosmological perturbations with the Schr\"odinger wave functional, and we generalize it to the case with interacting boundary terms. In Sect.~\ref{sec:boundary term}, we discuss the boundary term of $\zeta$ and its role on the well-poseness of variational principle, and its contribution to the decoherence rate is calculated. In Sect.~\ref{sec:divergences}, we discuss the UV and IR divergences, showing that the former is related to the renormalization of the wave functional with field redefinition, and the latter is associated with a real IR cutoff of the duration of inflation. In Sect.~\ref{sec:inflaton_self-interaction}, we study the decoherence in the semi-classical gravity in which the inflatons can only interact through self-interactions, and we discuss the difference between the cases with and without the non-classical gravity. Sect.~\ref{sec:conclusion} is our conclusion and outlook.

We set some notations for convenience. The system and environment are denoted by $\xi$ and $\mathcal{E}$ respectively, and their momentum indexes are $\bf q$ and $\bf k$ respectively. The integral over two environment modes with a system mode with momentum conservation is denoted by $\int_{\bf {k}+\bf{k}'=-\bf{q}}=\int\frac{d^3k}{(2\pi)^3}\frac{d^3 k'}{(2\pi)^3}(2\pi)^3\delta^3(\bf {k}+\bf{k}'+\bf{q})$, and we denote $\int_{{\bf {k}}_1,\cdots,{\bf {k}}_n,{\bf q}}=\int\frac{d^3k_1}{(2\pi)^3}\cdots\frac{d^3k_n}{(2\pi)^3}\frac{d^3q}{(2\pi)^3}(2\pi)^3\delta^3({\bf k}_1+\cdots+{\bf k}_n+\bf{q})$ as summing over all momentum-conserving modes. 

We also define the system and environment used in this paper. For the modes of cosmological perturbations observable with the comoving momenta within $\{\bf q\}$, we consider them as the system, whereas other modes with comoving momenta $\{\bf k\}$ are considered as the environment. A common choice is to separate the two parts with the horizon ${\rm Max} (q)={\rm Min} (k)=a(\tau)H$ with $\tau$ as the conformal time, but this can also include some unobserved super-horizon modes in the environment. We will show that the super-horizon environment dominates the contribution of decoherence, as shown in Sect.~\ref{sec:divergences} and \ref{sec:inflaton_self-interaction} for $\zeta$'s boundary term and inflaton's self-interaction respectively.

\section{Decoherence rate with the Schr\"odinger wave functional}
\label{sec:wave functional}
\subsection{The formalism}
We review the formalism in \cite{Nelson:2016kjm} to calculate the decoherence rate. For a weak cubic interaction of $\zeta$, we expect the couplings between a system mode and two environment modes in the momentum space
\begin{align}
H_{\rm int}(\tau)\supset \int_{{\bf k},{\bf k}',{\bf q}} \tilde{H}^{\rm int}_{\bf{k},\bf{k}',\bf{q}}(\tau)\mathcal{E}_{\bf k}\mathcal{E}_{{\bf k}'}\xi_{\bf q} \ , \label{eq:int_Hamiltonian}
\end{align}
and we can make a corresponding ansatz of the wave functional
\begin{align}
\langle\mathcal{E},\xi|\Psi(\tau)\rangle=\exp\left(\int_{{\bf k},{\bf k}',{\bf q}}  \mathcal{F}_{{\bf k},{\bf k}',{\bf q}}\mathcal{E}_{{\bf k}}\mathcal{E}_{{\bf k}'}\xi_{{\bf q}}\right)\Psi_G(\mathcal{E},\xi) \ . \label{eq:wave functional_ansatz}
\end{align}
The Gaussian part is separable \footnote{In \cite{Nelson:2016kjm,Burgess:2014eoa}, the Gaussian factor is defined by $\exp\left(-\int \frac{d^3p}{(2\pi)^3}A_p\zeta_{\bf p}\zeta_{\bf p}^*\right)$, and the conjugate momentum $-i\frac{\delta}{\delta \zeta_{\bf p}^*}$ operates once to the $\zeta_{\bf p}^*$, independent to $\zeta_{\bf p}$. However, here the conjugated momentum defined by $-i\frac{\delta}{\delta \zeta_{-{\bf p}}}$ operators both in $\zeta_{\pm{\bf p}}$, so there is $\frac{1}{2}$ factor to compensate this. Similar convention is also used in, e.g. \cite{Baumann:2020dch}.}
\begin{align}
\Psi_G(\mathcal{E},\xi)&=\Psi_G^{(\mathcal{E})}(\mathcal{E})\Psi_G^{(\xi)}(\xi) \nonumber \\
&=N_G(\tau)\exp\left(-\frac{1}{2}\int\frac{d^3k}{(2\pi)^3}A_k(\tau) \mathcal{E}_{\bf k}\mathcal{E}_{-{\bf k}}-\frac{1}{2}\int\frac{d^3q}{(2\pi)^3}A_q(\tau) \xi_{\bf q}\xi_{-{\bf q}}\right) \ ,
\end{align}
where $N_G=N_G^{(\mathcal{E})}N_G^{(\xi)}=\left(\prod_{\bf k}N_{G,{\bf k}}^{(\mathcal{E})}\right)\left(\prod_{\bf q}N_{G,{\bf q}}^{(\xi)}\right)$ is the separable normalization factor, and
\begin{align}
A^{(\zeta)}_p(\tau)=2p^3\frac{\epsilon M_p^2}{H^2}\frac{1-\frac{i}{p\tau}}{1+p^2\tau^2} \ ,
\end{align}
for the comoving curvature perturbation $\zeta$, whereas 
\begin{align}
A^{(\varphi)}_p(\tau)=-\frac{i}{(H\tau)^2} \frac{u'_p(\tau)}{u_p(\tau)} \ ,
\end{align}
for the inflaton perturbation $\varphi$, where the mode function
\begin{align}
u_p(\tau)=-e^{\frac{i\nu\pi}{2}-\frac{i\pi}{4}}\frac{\sqrt{\pi}}{2} \frac{H}{\sqrt{p^3}}(-p\tau)^{\frac{3}{2}}H_\nu^{(2)}(-p\tau) \ ,
\end{align}
with ${\bf p}$ includes both system and environment modes, and 
\begin{align}
\nu^2&=\frac{9}{4}-\frac{V''-2H^2(3\epsilon-\epsilon^2+\epsilon\eta)}{H^2} \ .
\end{align}
It is noteworthy the choice of $H^{(2)}_{\nu}(-p\tau)$, instead of $H^{(1)}_{\nu}(-p\tau)$, keeps the right sign of the exponent of Gaussian wave functional in the free theory. The power spectrum is evaluated through the variance
\begin{align}
\langle \mathcal{E}_{\bf k}\mathcal{E}_{{\bf k}'}\rangle&=(2\pi)^3\delta^3({\bf k}+{\bf k}')P_k(\tau) \nonumber \\
\langle \xi_{\bf q}\xi_{{\bf q}'}\rangle&=(2\pi)^3\delta^3({\bf q}+{\bf q}')P_q(\tau) \ ,
\end{align}
with 
\begin{align}
P_p(\tau)&=\frac{1}{2{\rm Re}A_p(\tau)} \ .
\end{align}

By solving the Schr\"odinger equation with the interaction (\ref{eq:int_Hamiltonian}) and ignoring the terms of $\mathcal{O}\left((\tilde{H}^{\rm int})^2\right)$ or beyond the cubic non-Gaussianity, the exponent of non-Gaussian part is
\begin{align}
\mathcal{F}_{\bf{k},\bf{k}',\bf{q}}\approx i\int^\tau_{\tau_i}\frac{d\tau'}{H\tau'}\tilde{H}^{\rm int}_{\bf{k},\bf{k}',\bf{q}}(\tau')\frac{u_k(\tau')u_{k'}(\tau')u_q(\tau')}{u_k(\tau)u_{k'}(\tau)u_q(\tau)}+\mathcal{O}\left((\tilde{H}^{\rm int})^2\right) \ , \label{eq:F_definition}
\end{align}
and an alternative derivation using the unitary operator is shown in \cite{Liu:2016aaf}. With the ansatz (\ref{eq:wave functional_ansatz}), the reduced density matrix for the system $\xi$ is obtained by integrating out the environment $\mathcal{E}$
\begin{align}
\rho_R(\xi,\tilde{\xi})&=\Psi_G^{(\xi)}(\xi)\left[\Psi_G^{(\xi)}(\tilde\xi)\right]^* \left\langle \exp\left(X\right)\right\rangle_{\mathcal{E}} \ , \label{eq:reduced_density_matrix}
\end{align}
where
\begin{align}
X&\equiv\int_{{\bf k},{\bf k}',{\bf q}}\mathcal{E}_{\bf k}\mathcal{E}_{{\bf k}'}\left(\xi_{\bf q}\mathcal{F}_{{\bf k},{\bf k}',{\bf q}}+\tilde\xi_{\bf q}\mathcal{F}^*_{{\bf k},{\bf k}',{\bf q}}\right) \ , \label{eq:X_def}
\end{align}
and
\begin{align}
\langle \cdots \rangle_{\mathcal{E}}\equiv\int D\mathcal{E}\left|\Psi_G^{(\mathcal{E})}\right|^2 ( \cdots ) \ .
\end{align}
The decaying exponent of the off-diagonal terms of $\rho_R(\xi,\tilde{\xi})$ quantifies the decoherence effect, and thus we define the decoherence factor as the absolute ratio of off-diagonal to diagonal terms
\begin{align}
D(\xi,\tilde{\xi})\equiv\left| \frac{\rho_R(\xi,\tilde{\xi})}{\sqrt{\rho_R(\xi,\xi)\rho_R(\tilde{\xi},\tilde{\xi})}}\right| \ .
\end{align}
The leading $\mathcal{O}(|\mathcal{F}|^2)$ contribution of the exponent of $D(\xi,\tilde{\xi})$ is approximated by the variance of $X$ defined in (\ref{eq:X_def}), and thus the decoherence factor for a system mode with comoving momentum $\bf q$ is
\begin{align}
D(\xi_{\bf q},\tilde\xi_{\bf q})\approx\exp\left[-\frac{|\xi_{\bf q}-\tilde\xi_{\bf q}|^2}{V}\int_{{\bf k}+{\bf k}'=-{\bf q}}P_k(\tau)P_{k'}(\tau)\left|\mathcal{F}_{{\bf k},{\bf k}',{\bf q}}\right|^2+\mathcal{O}(|\mathcal{F}|^4)\right] \ , \label{eq:deco_factor}
\end{align}
where the volume $V=(2\pi)^3\delta^3({\bf 0})$ is introduced for the discretization of momentum space
\begin{align}
\int \frac{d^3q}{(2\pi)^3}\to\frac{\sum_{\bf q}}{V} \ .
\end{align}
It is noteworthy that we use the absolute value of $\mathcal{F}_{\bf{k},\bf{k}',\bf{q}}$ in the integrand of (\ref{eq:deco_factor}) (see \cite{Liu:2016aaf} for the derivation), whereas \cite{Nelson:2016kjm} ignores the contribution from the real part by arguing it is small, but here we do not make such an assumption. 

The difference the two field configurations in (\ref{eq:deco_factor}) can be approximated by its expectation value $\langle |\xi_{\bf q}-\tilde\xi_{\bf q}|^2 \rangle=2 P_qV$, and the decoherence rate is defined by the expectation value of the minus exponent \footnote{We notice that the coefficient of (\ref{eq:Gamma_deco}), $4\pi^2$, disagrees with the one in \cite{Nelson:2016kjm} which is $8\pi$. Such a coefficient comes from using the dimensionless power spectrum $P_q=\frac{2\pi^2}{q^3}\Delta^2$, so we expect that the result is proportional to $\pi^2$, but the difference of coefficient does not change the conclusion in this paper.}
\begin{align}
\Gamma(\tau)=\frac{4\pi^2 \Delta^2}{q^3}\int_{\bf {k}+\bf{k}'=-\bf{q}}P_k(\tau)P_{k'}(\tau)|\mathcal{F}_{\bf{k},\bf{k}',\bf{q}}(\tau)|^2 \ , \label{eq:Gamma_deco}
\end{align}
where $\Delta^2$ is the late-time value of $\frac{q^3}{2\pi^2}P_q$. Although $\Gamma$ is called the decoherence rate, following the name in \cite{Nelson:2016kjm}, it is a dimensionless quantity, and decoherence happens when $\Gamma(\tau)\sim \mathcal{O}(1)$.

\subsection{With interacting boundary terms}
Now we generalize the formalism to the case with an interacting boundary term involving $\zeta$ and independent to $\dot{\zeta}$
\begin{align}
H_{\rm bd}(\zeta,t)&=\partial_t K(\zeta,t) \ ,
\end{align}
and the corresponding unitary operator is
\begin{align}
U(\tau,\tau_0)&=U_0(\tau,\tau_0)T \exp\left[-i\int^t_{t_0}dt_1H^I_{\rm bd}(t_1)\right] \nonumber  \\
&=U_0(\tau,\tau_0)\sum_{n=0}^\infty(-i)^n\int^t_{t_0}dt_1\int^{t_1}_{t_0}dt_2\cdots\int^{t_{n-1}}_{t_0}dt_n\partial_{t_1}K_I\partial_{t_2}K_I\cdots \partial_{t_n}K_I \nonumber \\
&=U_0(\tau,\tau_0)\sum_{n=0}^\infty \frac{(-i)^n}{n!}K_I^n(t) \nonumber \\
&=\sum_{n=0}^\infty \frac{(-i)^n}{n!}\left[U_0(\tau,\tau_0)K_I(t)U_0^{-1}(\tau,\tau_0)\right]^nU_0(\tau,\tau_0) \nonumber \\
&=\exp\left[-i K_S(\zeta,t)\right]U_0(\tau,\tau_0) \ , \label{eq:boundary_evolution}
\end{align}
where the labels $I$ and $S$ denote the interaction and Schr\"odinger pictures respectively. Therefore, the boundary term contributes a general non-Gaussian phase to the wave functional
\begin{align}
\langle\zeta|\Psi(\tau)\rangle&=\exp\left[-iK(\zeta,\tau)\right]\Psi_G(\zeta) \ , \label{eq:wave functional_K_phase}
\end{align}
where the label $S$ is dropped as there is no ambiguity. In the case with a cubic boundary term $K(\zeta,\tau)\propto \zeta^3$, the solution fits the ansatz (\ref{eq:wave functional_ansatz}) exactly with an purely imaginary $\mathcal{F}$ after separating the system $\xi$ and environment $\mathcal{E}$. In Sect.~\ref{sec:boundary term}, we will use (\ref{eq:wave functional_K_phase}) as a starting point to calculate the decoherence rate by the boundary term.

\section{Boundary term of $\zeta$ and decoherence}
\label{sec:boundary term}
The dominated boundary term for super-horizon modes is \cite{Maldacena:2002vr} 
\begin{align}
\mathcal{L}&=\partial_t\left(-2a^3HM_p^2 e^{3\zeta}\right)+{\rm derivatives} \ . \label{eq:boundary_plus_derivatives}
\end{align}
In the classical level, such a total-derivative term cannot contribute any dynamics as it does not contribute to the Euler-Lagrange equation. In the quantum level, boundary terms independent to $\dot{\zeta}$ cause no effect to the correlations $\langle \zeta^n \rangle$, easily shown by canceling the related phase factors with the wave functional or the in-in path integral, whereas boundary terms depending on $\dot{\zeta}$ can be removed by field redefinitions \cite{Arroja:2011yj,Burrage:2011hd}. Because of these reasons, the boundary term (\ref{eq:boundary_plus_derivatives}) is usually disregarded in literature. However, we will show that the non-Gaussian phase of the wave functional from the boundary term (\ref{eq:boundary_evolution}) can affect the correlators involving $\dot{\zeta}$ and decoherence. In this section, we will first discuss the relation between the boundary term and the well-defined variational principle. We then provide two approaches to calculate the decoherence rate by such a term: perturbation and the saddle-point approximation. Finally, we will discuss the decoherence in $\delta\phi$-gauge and the importance of choosing the boundary hypersurface.

\subsection{The well-defined variational principle}
In this subsection, we will show that the boundary term in (\ref{eq:boundary_plus_derivatives}) is necessary to exist for the well-defined variational principle. We start with the ADM decomposition of the Ricci scalar \cite{Wald:1984rg,Wang:2013zva}
\begin{align}
R&= {^{(3)}R}-K^2+K^\mu_\nu K^\nu_\mu-2\nabla_\mu\left(-K n^\mu+n^\nu \nabla_\nu n^\mu\right) \nonumber \\
&=R_{\rm ADM}-2\nabla_\mu\left(-K n^\mu+n^\nu \nabla_\nu n^\mu\right) \ , \label{eq:decompose_Ricci}
\end{align}
where $n^\mu$ is the normal of the hypersurface chosen to decompose the spacetime, the extrinsic curvature
\begin{align}
K_{\mu\nu}=\left(\delta^\rho_\mu+n^\rho n_\mu\right)\nabla_\rho n_\nu \ ,
\end{align}
and $R_{\rm ADM}$ is the part kept to derive the Lagrangian of cosmological perturbations (\ref{eq:boundary_plus_derivatives}) under the ADM formalism, whereas the divergence term is usually neglected in literature. One may wonder whether keeping such a divergence term cancels the total-derivative term in (\ref{eq:boundary_plus_derivatives}), and its explicit form shows that this cannot
\begin{align}
-\sqrt{-g}M_p^2\nabla_\mu\left(-K n^\mu+n^\nu \nabla_\nu n^\mu\right)\approx \partial_t \left[3a^3H M_p^2e^{3\zeta}\right]+\mathcal{O}(\epsilon,\partial^2\zeta) \ , \label{eq:divergence_term}
\end{align}
where this is checked up to $\mathcal{O}(\zeta^3)$. Therefore, both the Einstein-Hilbert and ADM actions bring us the total-derivative term in (\ref{eq:boundary_plus_derivatives}).

Using the Einstein-Hilbert action alone is not a well-defined variation as the Ricci scalar includes second-order derivatives, and this can explain why the coefficient of the total-derivative term (\ref{eq:boundary_plus_derivatives}) is unique. To see this, we realize that the divergence term in (\ref{eq:decompose_Ricci}) has the same contribution as the Gibbons-Hawking-York (GHY) boundary term
\cite{York:1972sj,Gibbons:1976ue,York:1986lje}
\begin{align}
\int_{\mathcal{M}} d^4x \sqrt{-g}M_p^2\nabla_\mu\left(-K n^\mu+n^\nu \nabla_\nu n^\mu\right)&=-M_p^2\int_{\partial\mathcal{M}}d^3y \sqrt{h}K \nonumber \\
&=S_{\rm GHY} \ , \label{eq:GHY}
\end{align}
where $\mathcal{M}$ is the manifold, and $h_{ij}$ is the induced metric on the hypersurface $\mathcal{\partial\mathcal{M}}$, defined as the $t={\rm const}$ hypersurface in the $\zeta$-gauge. For making the variation well-defined, we may either start with $R_{\rm ADM}$ or the Einstein-Hilbert action plus the GHY boundary term, and the results are identical to (\ref{eq:boundary_plus_derivatives}) which includes the total-derivative term:
\begin{align}
\int_{\mathcal{M}} d^4x \sqrt{-g} \left[\frac{M_p^2}{2}R_{\rm ADM}+\mathcal{L}_{\phi}\right]&=\int_{\mathcal{M}} d^4x \sqrt{-g} \left[\frac{M_p^2}{2}R+\mathcal{L}_{\phi}\right]+S_{\rm GHY} \nonumber \\
&=\int dt \  d^3 x \ \partial_t\left(-2a^3HM_p^2 e^{3\zeta}\right)+{\rm derivatives} \ . \label{eq:ADM_vs_EH_GHY}
\end{align}
As discussed in \cite{Chakraborty:2016yna}, the GHY boundary term is the only option to make the variation well-defined if the induced metric $h^{ij}$ is fixed on the hypersurface $\mathcal{\partial\mathcal{M}}$, implying that the boundary term is unique.

\subsection{Decoherence rate from perturbative expansion}
The full wave functional is
\begin{align}
\Psi(\zeta)&=N_G(\tau)\exp\left(-\frac{1}{2}\int\frac{d^3p}{(2\pi)^3}A_p(\tau) \zeta_{\bf p}\zeta_{-{\bf p}}\right)\exp\left(-2ia^3H M_p^2\int d^3x \ e^{3\zeta}\right) \nonumber  \\
&=N_G(\tau)\exp\left(-\frac{1}{2}\int\frac{d^3p}{(2\pi)^3}A_p(\tau) \zeta_{\bf p}\zeta_{-{\bf p}}\right)\exp\left(\frac{2}{27}\mathcal{F}\int d^3x\ e^{3\zeta}\right) \ , \label{eq:full_bd_wave functional}
\end{align} 
where the coefficient of $\mathcal{F}=-27ia^3HM_p^2$ is chosen to agree with the ansatz of cubic exponent (\ref{eq:wave functional_ansatz}), and we assume that this is small enough to be treated perturbatively in this subsection. For the system and environment separated by comoving momenta, we have
\begin{align}
\zeta(x)&=\int \frac{d^3k}{(2\pi)^3}\mathcal{E}_{\bf k}e^{i{\bf k}\cdot{\bf x}}+\int \frac{d^3q}{(2\pi)^3}\xi_{\bf q}e^{i{\bf q}\cdot{\bf x}} \ ,
\end{align}
so we write $\zeta= \mathcal{E} + \xi$ and $\tilde\zeta= \mathcal{E} + \tilde\xi$.
The decoherence factor is
\begin{align}
D(\xi,\tilde{\xi})&=\Big|\left\langle\exp\left[i\frac{2{\rm Im}\mathcal{F}}{27}\int d^3x\ \left(e^{3\zeta}-e^{3\tilde\zeta}\right)\right] \right\rangle_{\mathcal{E}}\Big| \nonumber \\
&=\Big|\left\langle e^{i\Delta S_{\rm bd}(\zeta,\tilde{\zeta})} \right\rangle_{\mathcal{E}} \Big| , \label{eq:action_difference}
\end{align} 
which is related to the action difference $\Delta S_{\rm bd}(\zeta,\tilde{\zeta})=S_{\rm bd}(\zeta)-S_{\rm bd}(\tilde\zeta)$ of two field configurations with integrating out the environment. 

\begin{figure}
\centering

\tikzset{every picture/.style={line width=0.75pt}} %set default line width to 0.75pt        

\begin{tikzpicture}[x=0.75pt,y=0.75pt,yscale=-1,xscale=1]
	
	%Straight Lines [id:da36441219715642403] 
	\draw [line width=1.5]    (26,80.23) -- (70.74,80.23) ;
	%Straight Lines [id:da7213817559419178] 
	\draw [color={rgb, 255:red, 255; green, 0; blue, 0 }  ,draw opacity=1 ][line width=1.5]    (70.74,80.23) -- (115.49,80.23) ;
	%Straight Lines [id:da7372382697710302] 
	\draw [line width=1.5]    (115.49,80.23) -- (160.23,80.23) ;
	%Straight Lines [id:da5361669254565937] 
	\draw [line width=1.5]    (183,80.23) -- (227.74,80.23) ;
	%Straight Lines [id:da17025225803750121] 
	\draw [line width=1.5]    (272.49,80.23) -- (317.23,80.23) ;
	%Shape: Circle [id:dp4717266950172867] 
	\draw  [color={rgb, 255:red, 255; green, 0; blue, 0 }  ,draw opacity=1 ][line width=1.5]  (227.74,80.23) .. controls (227.74,67.88) and (237.76,57.86) .. (250.12,57.86) .. controls (262.47,57.86) and (272.49,67.88) .. (272.49,80.23) .. controls (272.49,92.59) and (262.47,102.6) .. (250.12,102.6) .. controls (237.76,102.6) and (227.74,92.59) .. (227.74,80.23) -- cycle ;
	%Straight Lines [id:da9670882364057749] 
	\draw [line width=1.5]    (346,80.23) -- (390.74,80.23) ;
	%Straight Lines [id:da6391921827694815] 
	\draw [color={rgb, 255:red, 255; green, 0; blue, 0 }  ,draw opacity=1 ][line width=1.5]    (390.74,80.23) -- (435.49,80.23) ;
	%Straight Lines [id:da4739104720703793] 
	\draw [line width=1.5]    (435.49,80.23) -- (480.23,80.23) ;
	%Shape: Circle [id:dp8061027440197575] 
	\draw  [color={rgb, 255:red, 255; green, 0; blue, 0 }  ,draw opacity=1 ][line width=1.5]  (413.12,57.86) .. controls (413.12,45.5) and (423.13,35.49) .. (435.49,35.49) .. controls (447.84,35.49) and (457.86,45.5) .. (457.86,57.86) .. controls (457.86,70.22) and (447.84,80.23) .. (435.49,80.23) .. controls (423.13,80.23) and (413.12,70.22) .. (413.12,57.86) -- cycle ;
	%Straight Lines [id:da004549936214974126] 
	\draw [line width=1.5]    (504,80.23) -- (548.74,80.23) ;
	%Straight Lines [id:da7898933368775003] 
	\draw [line width=1.5]    (593.49,80.23) -- (638.23,80.23) ;
	%Shape: Circle [id:dp7615630672843692] 
	\draw  [color={rgb, 255:red, 255; green, 0; blue, 0 }  ,draw opacity=1 ][line width=1.5]  (548.74,80.23) .. controls (548.74,67.88) and (558.76,57.86) .. (571.12,57.86) .. controls (583.47,57.86) and (593.49,67.88) .. (593.49,80.23) .. controls (593.49,92.59) and (583.47,102.6) .. (571.12,102.6) .. controls (558.76,102.6) and (548.74,92.59) .. (548.74,80.23) -- cycle ;
	%Straight Lines [id:da6941407905848032] 
	\draw [color={rgb, 255:red, 255; green, 0; blue, 0 }  ,draw opacity=1 ][line width=1.5]    (548.74,80.23) -- (593.49,80.23) ;
	
	% Text Node
	\draw (41.77,84.31) node [anchor=north west][inner sep=0.75pt]    {$q$};
	% Text Node
	\draw (131.26,85.1) node [anchor=north west][inner sep=0.75pt]    {$q$};
	% Text Node
	\draw (83.62,61.7) node [anchor=north west][inner sep=0.75pt]  [color={rgb, 255:red, 255; green, 0; blue, 0 }  ,opacity=1 ]  {$P_{q}$};
	% Text Node
	\draw (198.77,84.31) node [anchor=north west][inner sep=0.75pt]    {$q$};
	% Text Node
	\draw (288.26,85.1) node [anchor=north west][inner sep=0.75pt]    {$q$};
	% Text Node
	\draw (240.62,39.7) node [anchor=north west][inner sep=0.75pt]  [color={rgb, 255:red, 255; green, 0; blue, 0 }  ,opacity=1 ]  {$P_{k}$};
	% Text Node
	\draw (241.62,105.7) node [anchor=north west][inner sep=0.75pt]  [color={rgb, 255:red, 255; green, 0; blue, 0 }  ,opacity=1 ]  {$P_{k'}$};
	% Text Node
	\draw (361.77,84.31) node [anchor=north west][inner sep=0.75pt]    {$q$};
	% Text Node
	\draw (451.26,84.1) node [anchor=north west][inner sep=0.75pt]    {$q$};
	% Text Node
	\draw (403.62,83.7) node [anchor=north west][inner sep=0.75pt]  [color={rgb, 255:red, 255; green, 0; blue, 0 }  ,opacity=1 ]  {$P_{q}$};
	% Text Node
	\draw (426.62,18.7) node [anchor=north west][inner sep=0.75pt]  [color={rgb, 255:red, 255; green, 0; blue, 0 }  ,opacity=1 ]  {$P_{k}$};
	% Text Node
	\draw (519.77,84.31) node [anchor=north west][inner sep=0.75pt]    {$q$};
	% Text Node
	\draw (609.26,85.1) node [anchor=north west][inner sep=0.75pt]    {$q$};
	% Text Node
	\draw (561.62,40.7) node [anchor=north west][inner sep=0.75pt]  [color={rgb, 255:red, 255; green, 0; blue, 0 }  ,opacity=1 ]  {$P_{k_{1}}$};
	% Text Node
	\draw (562.62,103.7) node [anchor=north west][inner sep=0.75pt]  [color={rgb, 255:red, 255; green, 0; blue, 0 }  ,opacity=1 ]  {$P_{k_{3}}$};
	% Text Node
	\draw (562.62,62.7) node [anchor=north west][inner sep=0.75pt]  [color={rgb, 255:red, 255; green, 0; blue, 0 }  ,opacity=1 ]  {$P_{k_{2}}$};

\end{tikzpicture}

\caption{The diagrams of the four terms in (\ref{eq:off_diagonal_all_order}).}
\label{fig:diagrams_four_terms}
\end{figure}
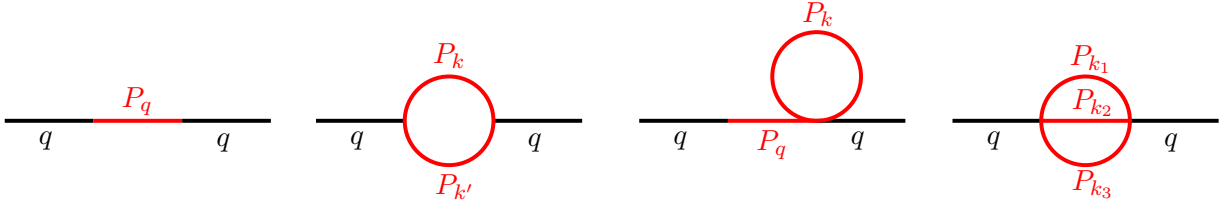

Expanding the action up to $\mathcal{O}(\zeta^4)$ and collecting the terms including at least one environment mode
\begin{align}
&i\Delta S_{\rm bd}(\zeta,\tilde{\zeta}) \nonumber \\
&\supset \frac{2i{\rm Im}\mathcal{F}}{27} \sum_{n=2}^\infty\frac{3^n}{n!}\sum_{m=1}^{n-1}\binom{n}{m}\int_{{\bf k}_1\cdots{\bf k}_m,{\bf q}_1\cdots{\bf q}_{n-m}}\mathcal{E}_{{\bf k}_1}\cdots \mathcal{E}_{{\bf k}_m}\left(\xi_{{\bf q}_1}\cdots\xi_{{\bf q}_{n-m}}-\tilde\xi_{{\bf q}_1}\cdots\tilde\xi_{{\bf q}_{n-m}}\right) \nonumber \\
&\approx
 i {\rm Im}\mathcal{F}\Bigg\{\frac{2}{3}\int_{\bf q}\mathcal{E}_{-{\bf q}}(\xi_{\bf{q}}-\tilde\xi_{\bf{q}})+\int_{\bf{k},\bf{k}',\bf{q}}\mathcal{E}_{\bf k}\mathcal{E}_{{\bf k}'}(\xi_{\bf{q}}-\tilde\xi_{\bf{q}})+\int_{\bf{k},\bf{q},\bf{q}'}\mathcal{E}_{\bf k}(\xi_{{\bf q}}\xi_{\bf{q}'}-\tilde\xi_{\bf{q}}\tilde\xi_{\bf{q}'}) \nonumber \\
&+\frac{1}{4}\Big[4\int_{\bf{k}_1,\bf{k}_2,\bf{k}_3,\bf{q}}\mathcal{E}_{{\bf k}_1}\mathcal{E}_{{\bf k}_2}\mathcal{E}_{{\bf k}_3}(\xi_{\bf{q}}-\tilde\xi_{\bf{q}})+6\int_{\bf{k}_1,\bf{k}_2,\bf{q}_1,\bf{q}_2}\mathcal{E}_{{\bf k}_1}\mathcal{E}_{{\bf k}_2}(\xi_{{\bf q}_1}\xi_{\bf{q}_2}-\tilde\xi_{\bf{q}_1}\tilde\xi_{\bf{q}_2}) \nonumber \\
&+4\int_{\bf{k},\bf{q}_1,\bf{q}_2,\bf{q}_3}\mathcal{E}_{{\bf k}}(\xi_{{\bf q}_1}\xi_{\bf{q}_2}\xi_{\bf{q}_3}-\tilde\xi_{\bf{q}_1}\tilde\xi_{\bf{q}_2}\tilde\xi_{\bf{q}_3})
\Big]
\Bigg\}+\mathcal{O}(\zeta^5) \ , \label{eq:off_diagonal_all_order}
\end{align}
and the leading order of decoherence factor is its variance
\begin{align}
D(\xi,\tilde{\xi})&\approx\exp\Bigg\{-\frac{\left\langle\left[\Delta S_{\rm bd}(\zeta,\tilde{\zeta})\right]^2\right\rangle_{\mathcal{E},c}}{2}+\mathcal{O}(|\mathcal{F}|^3)\Bigg\} \nonumber \\
&\approx\exp\Bigg[-\frac{2({\rm Im}\mathcal{F})^2}{9}\int_{\bf q}|\xi_{\bf q}-\tilde\xi_{\bf q}|^2 \Big(P_q+ \frac{9}{2}\int_{{\bf k}+{\bf k}'=-{\bf q}}P_k P_{k'}+9\int_{{\bf k}}P_kP_q \nonumber \\
&+\frac{27}{2}\int_{{\bf k}_1+{\bf k}_2+{\bf k}_3=-{\bf q}}P_{k_1}P_{k_2}P_{k_3}\Big)\Bigg]+\mathcal{O}(\xi^3) \ , \label{eq:deco_D_expand}
\end{align}
where $\langle\cdots\rangle_{\mathcal{E},c}$ means the connected part of correlation functions, and the terms with $\mathcal{O}(\xi^3)$ are neglected. The four terms in (\ref{eq:deco_D_expand}) can be described by diagrams, as shown in Fig. \ref{fig:diagrams_four_terms}.
It is clear that only the second and four terms, produced by the cubic and fourth orders of boundary term respectively, can contribute the decoherence, whereas the first and third terms cannot since the comoving momenta of environment modes cannot be equal to $q$. Consider the terms up to 1-loop \footnote{For the terms with more loops, they involve higher order of the power spectrum $P_k$, suppressed with $\frac{H^2}{\epsilon M_p^2}\sim \mathcal{O}(10^{-9})$ .}, the decoherence rate of the system mode ${\bf q}$ is 
\begin{align}
\Gamma_\zeta^{\rm bd}(\tau)\approx\frac{4\pi^2 \Delta^2_{\zeta}}{q^3}|\mathcal{F}|^2\int_{{\bf k}+{\bf k}'=-{\bf q}}P_k P_{k'} \ , \label{eq:deco_1loop_boundary}
\end{align}
which agrees with (\ref{eq:Gamma_deco}).

\subsection{Decoherence rate from the saddle-point approximation}
Since $\mathcal{F}=-27ia^3HM_p^2$ is not slow-roll suppressed, it may not be obvious why it can act as a perturbative parameter. Here we also apply the saddle-point approximation to calculate the functional integral of (\ref{eq:action_difference}), and we start with the action difference
\begin{align}
i\Delta S_{\rm bd}(\zeta,\tilde{\zeta}) &=\frac{2i{\rm Im}\mathcal{F}}{27}\int d^3x \ (e^{3\xi}-e^{3\tilde\xi})e^{3\mathcal{E}}\nonumber \\
&\approx\frac{2i{\rm Im}\mathcal{F}}{9}\int d^3x \ (\xi-\tilde{\xi})\sum_{n=0}^\infty\frac{3^n}{n!}\mathcal{E}^n+\mathcal{O}(\xi^2) \nonumber \\
&=\frac{2i{\rm Im}\mathcal{F}}{9}\left[\int d^3x \ (\xi-\tilde{\xi})\sum_{n\neq 1}^\infty\frac{3^n}{n!}\mathcal{E}^n+3\int \frac{d^3q}{(2\pi)^3}(\xi_{\bf q}-\tilde{\xi}_{\bf q})\mathcal{E}_{-{\bf q}}\right]+\mathcal{O}(\xi^2) \nonumber \\
&=\frac{2i{\rm Im}\mathcal{F}}{9}\int d^3x \ \Delta\xi \sum_{n\neq 1}^\infty\frac{3^n}{n!}\mathcal{E}^n+\mathcal{O}(\xi^2) \ , 
\end{align}
where the bilinear term vanishes since the system and environment modes do not have the same comoving momentum, and here the expansion relies on small $\xi$ but not small $|\mathcal{F}|$. With this, the functional integral (\ref{eq:action_difference}) has an integrand with phase starting from the order $\mathcal{E}^2$
\begin{align}
\Big|\left\langle e^{i\Delta S_{\rm bd}(\zeta,\tilde{\zeta})} \right\rangle_{\mathcal{E}} \Big|&\approx \Bigg| \left(N_G^{(\mathcal{E})}\right)^2 \int D\mathcal{E} \exp \Bigg[\frac{2i{\rm Im}\mathcal{F}}{9}\int d^3x \ \Delta\xi\sum_{n=2}^\infty\frac{3^n}{n!}\mathcal{E}^n \nonumber \\
& -\int d^3x d^3y \ \mathcal{E}({\bf x}) {\rm Re}A_{\mathcal{E}}(|{\bf x}-{\bf y}|)\mathcal{E}({\bf y}) \Bigg]\Bigg| 
 \ , \label{eq:functional_integral_explicit}
\end{align}
where ${\rm Re}A_\mathcal{E}(|{\bf x}-{\bf y}|)$ is the inverse Fourier transform of ${\rm Re}A_k$ which involves only the environment modes, and clearly the phase has a saddle point at $\mathcal{E}({\bf x})=0$. By applying the saddle-point approximation, (\ref{eq:functional_integral_explicit}) can be evaluated with the functional determinant of quadratic terms, and in the comoving-momentum space it has the form
\begin{align}
\Big|\left\langle e^{i\Delta S_{\rm bd}(\zeta,\tilde{\zeta})} \right\rangle_{\mathcal{E}} \Big| &\approx \Bigg| \left(N_G^{(\mathcal{E})}\right)^2 \int D\mathcal{E} \exp \left(i{\rm Im}\mathcal{F} \int_{{\bf k},-{\bf k}',{\bf q}} \mathcal{E}_{\bf k}\mathcal{E}_{-{\bf k}'}\Delta\xi_{\bf q}- \int \frac{d^3k}{(2\pi)^3} {\rm Re}A_k \mathcal{E}_{\bf k}\mathcal{E}_{-{\bf k}}\right)\Bigg| \nonumber \\
&=\Bigg|\prod_{{\bf k},{\bf k}'}\sqrt{\left(N_{G,{\bf k}}^{(\mathcal{E})}N_{G,{\bf k}' }^{(\mathcal{E})}\right)^2 \int D\mathcal{E}_{\bf k}D\mathcal{E}_{{\bf k}'}\exp\left(-M_{{\bf k},{\bf k}'}\mathcal{E}_{\bf k}\mathcal{E}_{-{\bf k}'}\right)}\Bigg| \nonumber \\
&=\left(\prod_{{\bf k},{\bf k}'}\frac{{\rm Re}A_k{\rm Re}A_{k'}}{V^2\det M_{{\bf k},{\bf k}'}}\right)^{1/4} \ ,
\end{align}
where the matrix is
\begin{align}
M_{{\bf k},{\bf k}'}&=\begin{pmatrix}
\frac{1}{V}{\rm Re}A_k & -\frac{i}{V^2}{\rm Im}\mathcal{F}\Delta \xi_{-({\bf k}-{\bf k}')} \\
-\frac{i}{V^2}{\rm Im}\mathcal{F}\Delta \xi_{{\bf k}-{\bf k}'}&\frac{1}{V}{\rm Re}A_{k'}
\end{pmatrix} \ ,
\end{align}
and the square root in the second line compensates the double counting of modes. After evaluating the functional determinant, the decoherence factor is
\begin{align}
D(\xi,\tilde{\xi})&=
\left(\prod_{{\bf k},{\bf k}'}\frac{1}{1+\frac{1}{V^2}\frac{({\rm Im}\mathcal{F})^2|\Delta \xi_{{\bf k}-{\bf k}'}|^2}{{\rm Re}A_k{\rm Re}A_{k'}}}\right)^{1/4} \nonumber \\
&\approx\exp\left[-\frac{({\rm Im}\mathcal{F})^2}{4}\frac{\sum_{{\bf k},{\bf k}'}}{V^2}\frac{|\Delta \xi_{{\bf k}-{\bf k}'}|^2}{{\rm Re}A_k{\rm Re}A_{k'}}\right] \nonumber \\
&= \exp\left[-|\mathcal{F}|^2\int_{{\bf k},{\bf k}',{\bf q}}P_kP_{k'}|\Delta \xi_{\bf q}|^2 \right] \ , \label{eq:decoherence_factor_D}
\end{align}
where we changed the index $-{\bf k}'\to{\bf k}'$ in the last line, and after selecting a system mode ${\bf q}$ the decoherence rate agrees with the result by the perturbation method (\ref{eq:deco_1loop_boundary}), showing that the result is independent to small $|\mathcal{F}|$.

\subsection{The boundary term with the $\delta \phi$-gauge and the choice of boundary}
\label{sec:deltaphi_gauge_boundary}
Before evaluating the decoherence rate explicitly, we discuss the boundary term in another choice of gauge, such as the spatially flat gauge (the $\delta\phi$-gauge) defined by letting $\zeta=0$. In such a gauge, the quantum degree of freedom is fully described by the inflaton fluctuation $\varphi$, and it is also a natural choice in the semi-classical limit where the metric fluctuation is classical. Starting with the ADM action defined in the LHS of (\ref{eq:ADM_vs_EH_GHY}), the action of $\varphi$ up to the cubic order is well-known in \cite{Maldacena:2002vr}, and here we focus on the boundary terms produced by taking integration by parts. The boundary terms of quadratic and cubic orders are
\begin{align}
\mathcal{L}_{{\rm Ibp},\varphi}&=M_p^2\partial_{t_\varphi}\left(-\frac{1}{2}a^3\epsilon H\varphi^2+\frac{a^3H \epsilon^{3/2}\varphi^3}{6\sqrt{2}}\right)\nonumber \\
&\approx M_p^2\partial_{t_\varphi}\left[-\frac{1}{3} a^3 \zeta ^2 H \epsilon ^2 (3+\epsilon \zeta)\right] \ , \label{eq:phi_boundary_from_IBP}
\end{align}
where we applied the field transformation
\begin{align}
\varphi \approx -\frac{\dot{\phi}}{H}\zeta+\mathcal{O}(\zeta^2) \ ,
\end{align}
for estimating the size of (\ref{eq:phi_boundary_from_IBP}), and the $t_\varphi$ is the time coordinate in the $\delta\phi$-gauge which is related to the time in the $\zeta$-gauge 
\begin{align}
t_\varphi&\approx t_\zeta-\frac{\varphi}{\dot{\phi}}+\mathcal{O}(\varphi^2) \nonumber \\
&\approx  t_\zeta+\frac{\zeta}{H}+\mathcal{O}(\zeta^2) \ . \label{eq:time_transform}
\end{align}

Clearly, (\ref{eq:phi_boundary_from_IBP}) is slow-roll suppressed, which is much smaller than the boundary term of $\zeta$-gauge (\ref{eq:ADM_vs_EH_GHY}), and naively we may conclude that the decoherence rate in the $\delta\phi$-gauge is much smaller. However, the coordinate transformation (\ref{eq:time_transform}) implies two different hypersurfaces defined by constant time, $t_\varphi={\rm const}$ and $t_\zeta={\rm const}$, and thus we cannot directly compare the boundary terms in different gauges which correspond to different physical boundaries. By constructing gauge-invariant actions order by order, it can be shown that different natural hypersurfaces defined in various gauges lead to different boundary terms \cite{Rigopoulos:2011eq,Prokopec:2012ug}. In the following calculation, we choose the boundary $\partial\mathcal{M}$ fixed by $t_\zeta=t$ since it is natural for observables frozen outside the horizon, and it is defined by position-dependent times in the $\delta\phi$-gauge, parameterized by 
\begin{align}
 t_\varphi({\bf y})=t+\frac{\zeta(t,{\bf y})}{H}\ , \ {\bf x}={\bf y} \ .  \label{eq:parameterization}
\end{align}
With this non-trivial boundary, we would like to see how this affects the Lagrangian, and it is easier to use the approach with the Einstein-Hilbert action and the GHY boundary term. Firstly, we calculate some geometric quantities on the boundary: the normal vector from the constraint (\ref{eq:parameterization}) is
\begin{align}
n^\mu=\frac{1}{\sqrt{(1-\epsilon\zeta)^2-\frac{1}{a^2H^2}\partial_i\zeta\partial_i\zeta}}\left(1-\epsilon\zeta,\frac{\partial_i\zeta}{a^2H}\right) \ ,
\end{align}
and the corresponding extrinsic curvature
\begin{align}
K&=\nabla_\mu n^\mu \nonumber \\
&\approx 3H +\mathcal{O}\left(\epsilon,\frac{\partial^2\zeta}{a^2}\right) \ ,
\end{align} 
where we keep the terms not slow-roll and super-horizon suppressed.
The induced metric is
\begin{align}
h_{ij}&=g_{\mu\nu}\frac{\partial x^\mu}{\partial y^i}\frac{\partial x^\nu}{\partial y^j}\Big|_{\partial\mathcal{M}} \nonumber \\
&=a(t_\varphi({\bf y}))^3\delta_{ij}-\frac{\partial_i\zeta\partial_j\zeta}{H^2} \ ,
\end{align} and the square root of its determinant
\begin{align}
\sqrt{h}&\approx a(t_\varphi({\bf y}))^3 +\mathcal{O}\left((\partial\zeta)^2\right) \nonumber \\
&=a(t)^3e^{3\zeta(t,{\bf y})} \ ,
\end{align}
and thus the GHY boundary term in the $\delta\phi$-gauge (\ref{eq:GHY}) is
\begin{align}
S_{\rm GHY}&\approx-M_p^2\int_{\partial \mathcal{M}}d^3y \ 3a(t)^3H e^{3\zeta(t,{\bf y})} +\mathcal{O}(\epsilon)  \ , \label{eq:GHY_delta_phi}
\end{align}
which agrees with the one in the $\zeta$-gauge (\ref{eq:divergence_term}). Not only does the Einstein-Hilbert action need a boundary term to make the variation well-defined, but the inflaton Lagrangian also needs one since the variation of $\phi$ on the boundary is not vanishing
\begin{align}
\delta\phi|_{\partial\mathcal{M}}&=\varphi(t,{\bf y}) \nonumber \\
&\approx -\frac{\dot{\phi}}{H}\zeta(t,{\bf y})+\mathcal{O}(\zeta^2) \ .
\end{align} 
We now study the variation of the matter action with a boundary term depending on $\phi$ and with negligible $\dot{\phi}$ dependence
\begin{align}
S_m&=\int_{\mathcal{M}} d^4x \sqrt{-g}\left[-\frac{1}{2}g^{\mu\nu}\partial_\mu\phi\partial_\nu\phi-V(\phi)\right]+\int_{\partial\mathcal{M}}d^3y \sqrt{h}\mathcal{L}_{{\rm bd},\phi}(\phi) \nonumber \\
&\approx \int d^3x\int dt_{\varphi} \ a^3\left[\frac{1}{2}{\dot{\phi}}^2-V(\phi)\right]+\int_{\partial\mathcal{M}}d^3y \sqrt{h}\mathcal{L}_{{\rm bd},\phi}(\phi) +\mathcal{O}(\varphi^2) \ ,
\end{align}
where we consider the background level in the last line. Varying the matter action gives
\begin{align}
\delta S_{m}&=-\int d^3x\int dt_{\varphi} \ a^3\left[\ddot{\phi}+3H\dot{\phi}+V'(\phi)\right]\delta\phi+\int_{\partial \mathcal{M}}d^3y \sqrt{h}\left(\dot{\phi}+\frac{\partial \mathcal{L}_{{\rm bd},\phi}}{\partial \phi}\right)\varphi \nonumber \\
&=0 \ ,
\end{align}
where making the first term zero gives the equation of motion, and the second term requires
\begin{align}
\frac{\partial \mathcal{L}_{{\rm bd},\phi}}{\partial\phi}&=-\dot{\phi} \nonumber \\
&\approx \frac{V'(\phi)}{3H}+\mathcal{O}(\epsilon,\eta) \ ,
\end{align}
where the last line comes from the equation of motion in the slow-roll limit, so we can choose 
\begin{align}
\mathcal{L}_{{\rm bd},\phi}&=\frac{V(\phi)}{3H} \nonumber \\
&\approx M_p^2H+\mathcal{O}(\epsilon,\eta) \ ,
\end{align}
which neglects any constant independent to the inflaton. The boundary action is thus
\begin{align}
\int_{\partial\mathcal{M}}d^3y \sqrt{h}\mathcal{L}_{{\rm bd},\phi}(\phi)\approx M_p^2 \int_{\partial\mathcal{M}}d^3y \ a(t)^3H e^{3\zeta(t,{\bf y})}+\mathcal{O}(\epsilon,\eta) \ ,
\end{align}
and adding the GHY boundary term (\ref{eq:GHY_delta_phi}) gives the same total boundary term of the $\zeta$-gauge (\ref{eq:ADM_vs_EH_GHY}).

\section{The IR and UV divergences of decoherence rate}
\label{sec:divergences}
\subsection{Resolve the IR divergence with the duration of inflation}
The IR divergence happens when some superhorizon modes are included in the environment, and their comoving momenta are too close to the one of the system. For the integral of decoherence rate \footnote{Here we also use another notation $P_\zeta(k)=P_k$ to represent the power spectrum of $\zeta$ since the argument of comoving momentum expressed in coordinates can be complicated.}
\begin{align}
\Gamma_\zeta^{\rm bd}&=\frac{4\pi^2\Delta^2_\zeta}{q^3}|\mathcal{F}|^2\int\frac{d^3k}{(2\pi)^3}P_\zeta (k) P_\zeta (|{\bf k}-{\bf q}|) \nonumber \\
&=\frac{4\pi^2\Delta^2_\zeta}{q^3}|\mathcal{F}|^2\frac{1}{4\pi^2}\int dk \ \frac{H^4 \left(k^2 \tau ^2+1\right) }{16 k^2 M_p^4 q \epsilon ^2}\left[\frac{1-\tau ^2 (k-q)^2}{| k-q| }+\frac{\tau ^2 (k+q)^2-1}{k+q}\right] \ , \label{eq:Gammadeco_boundary_divergence}
\end{align}
which has logarithmic divergence when $k\to 0$ or $k\to q$ ($k'\to 0$ due to conservation of comoving momenta). The logarithmic IR divergence is resolved by choosing a real cutoff $\log\left(\frac{q}{k_{\rm min}}\right)=N_q-N_{\rm IR}$, the e-folds from the beginning of inflation to the horizon crossing of the system mode $\bf q$, and this is finite as the duration of inflation cannot be infinite. Such an
IR cutoff has also been applied in \cite{Martin:2018zbe} to resolve the IR divergence in cosmological decoherence problem, and we will also choose their value $N_q-N_{\rm IR}\approx 10^4$ to estimate some results.

With the IR cutoff and excluding the environment modes indistinguishable from the system ($|k-q|, |k'-q|\leq k_{\rm min}$), (\ref{eq:Gammadeco_boundary_divergence}) is rewritten into \footnote{The prefactor of decoherence rate should be 729 instead of 81, and in the first two versions of the paper we had made a mistake to use $\mathcal{F}=-9ia^3HM_p^2$ where the correct prefactor should be 27.}
\begin{align}
&\Gamma_\zeta^{\rm bd}(q,\tau) \nonumber \\
&=\frac{4\pi^2\Delta^2_\zeta}{q^3}|\mathcal{F}|^2 \frac{1}{4\pi^2}\left(\int^{a\Lambda}_{k_{\rm min}, \ |k-q|>k_{\rm min}} dk \int^1_{-1}dx-\int^{2q+k_{\rm min}}_{k_{\rm min}, \ |k-q|>k_{\rm min}}dk \int^{\frac{k^2+q^2-(q-k_{\rm min})^2}{2kq}}_{\frac{k^2+q^2-(q+k_{\rm min})^2}{2kq}} dx\right) \nonumber \\
& \times k^2 P_\zeta (k) P_\zeta \left(\sqrt{k^2+q^2-2kqx}\right) \nonumber \\
&\approx \frac{729 \Delta_{\zeta} ^2}{16  \epsilon ^2}\left[\frac{2 \Lambda  }{H}\left(\frac{aH}{q}\right)^3+4 \left(\frac{aH}{q}\right)^6 \left(\Delta N-1\right)+4 \left(\frac{aH}{q}\right)^4 \Delta N-\left(\frac{aH}{q}\right)^2\right] \nonumber \\
& +\mathcal{O}\left(\frac{k_{\rm min}}{q}\right) \ , \label{eq:Gamma_IR_reg}
\end{align}
where $\Delta N=N_q-N_{\rm IR}$ and $\Lambda$ is a UV cutoff. The value of $\Delta N$ should be model-dependent, but a general minimum is determined by the condition if $\mathcal{O}\left(\frac{k_{\rm min}}{q}\right)$ is small, and this gives at least $\Delta N\gtrsim 2$.

\subsection{Resolve the UV divergence by renormalizing the wave functional by field redefinition}
Before resolving the UV divergence of decoherence rate, it is noteworthy that similar divergence also appears in the correlator $\langle \dot{\zeta} \dot{\zeta} \rangle$ \footnote{In the first two versions of the paper, we proposed to resolve the UV divergence of decoherence rate by renormalizing $\langle \dot{\zeta} \dot{\zeta} \rangle$, but renormalizing the wave functional is directly related to decoherence effect, and thus it is a more appropriate treatment.}. Since the 1-loop result (\ref{eq:deco_1loop_boundary}) depends only on the cubic phase of the wave functional (\ref{eq:full_bd_wave functional}), we can evaluate the following two-point function with functional derivatives
\begin{align}
\langle \dot{\zeta}_{-\bf{q}}\dot{\zeta}_{\bf{q}}\rangle'
&=\left(\frac{1}{2\epsilon M^2_p a^3}\right)^2\langle \pi_{-\bf{q}}\Psi|\pi_{-\bf{q}}\Psi\rangle' \nonumber \\
&=\left(\frac{1}{2\epsilon M^2_p a^3}\right)^2\int' D\zeta\left(\frac{\delta \Psi}{i\delta \zeta_{-{\bf q}}}\right)^*\left(\frac{\delta \Psi}{i\delta \zeta_{-{\bf q}}}\right) \nonumber \\
&=\left(\frac{1}{2\epsilon M^2_p a^3}\right)^2\left(\left|A_q\right|^2P_q+2\int_{\bf{k}+\bf{k}'=\bf{q}}\left|\mathcal{F}\right|^2P_kP_{k'}\right) \ , \label{eq:two_point_functional}
\end{align}
where $'$ means ignoring the momentum-conserving factor. To be familiar with the wave functional method, Appendix \ref{sec:appendix_compare_inin_schrodinger} compares the calculation of another correlator $\langle\dot{\zeta}_{{\bf k}_1}\dot{\zeta}_{{\bf k}_2}\dot{\zeta}_{{\bf k}_3}\rangle$ by the usual in-in formalism with the wave functional method. Comparing (\ref{eq:two_point_functional}) with the decoherence rate (\ref{eq:Gamma_deco}) shows that they share the same 1-loop result 
\begin{align}
\langle \dot{\zeta}_{-\bf{q}}\dot{\zeta}_{\bf{q}}\rangle'_{\rm 1-loop}&=
-\frac{729 H^5 }{128 M_p^4 \pi ^2 \epsilon ^4}\left[2  \Lambda  \tau ^3+\frac{H}{q^3} \left(2-2q^2\tau^2+q^4 \tau ^4\right)-\frac{4}{q^3} \left(H+H q^2 \tau ^2\right) \log
   \left(\frac{q}{k_{\rm min}}\right)\right] \ , \label{eq:1-loop_zetadot_zetadot}
\end{align}
and the UV-divergent term is proportional to $\delta^{3}(\bf{x})$ after Fourier transforming back to the real space, meaning that it is a contact term. We can also write the 1-loop correction in the real space
\begin{align}
\langle\dot\zeta({\bf x})\dot\zeta({\bf y})\rangle_{\rm 1-loop}&=\left(\frac{1}{2\epsilon M^2_p a^3}\right)^2  |\mathcal{F}|^2\langle\zeta^2({\bf x})\zeta^2({\bf y})\rangle \nonumber \\
&=-\frac{729H^5\tau^3}{64M_p^4\pi^2\epsilon^4}\Lambda\delta^3({\bf x}-{\bf y})+\cdots \ , \label{eq:two_pt_div}
\end{align}
where $\cdots$ are the terms independent to the UV-cutoff, and the composite operator in the first line with the form $\phi^2(x)\phi^2(y)$ has been discussed in some books \cite{Collins:1984xc,amit2005field}. 

To construct the counterterms for resolving the UV divergence, it is instructive to show that the Gaussian wave functional of $\zeta$ is related to the on-shell action (Hamiltonian-Jacobi functional) evaluated on the boundary $\partial \mathcal{M}$ at time $t$, as demonstrated in \cite{Maldacena:2002vr,Larsen:2003pf}. For the free action of $\zeta$, the on-shell action is evaluated with the integration by parts
\begin{align}
	S_2(\zeta)&=\int_{\partial M}d^3y \ \epsilon M_p^2 a^3 \dot{\zeta}\zeta+\int d^4x \ \frac{\zeta}{2}  \frac{\delta L_2}{\delta \zeta} \nonumber \\
	&=\int \frac{d^3p}{(2\pi)^3} \ \epsilon M_p^2 a^3 \frac{\dot{\zeta}_{\bf p}}{\zeta_{\bf p}} \zeta_{\bf p}\zeta_{-{\bf p}} \nonumber \\
	&=\frac{i}{2}\int \frac{d^3p}{(2\pi)^3} A_p\zeta_{\bf p}\zeta_{-{\bf p}}  \ , \label{eq:on-shell-action} 
\end{align}
where the time derivative $\dot{\zeta}_{\bf p}$ on $\partial \mathcal{M}$ is determined by the classical solution with the EOM $\frac{\delta L_2}{\delta \zeta}=0$, and the Gaussian wave functional is thus obtained by $\Psi(\zeta)\propto e^{i S_2(\zeta)}=e^{-\frac{1}{2}\int\frac{d^3p}{(2\pi)^3}A_p \zeta_{\bf p}\zeta_{-{\bf p}}}$. Recall that the power spectrum is related to the above coefficient of the wave functional
\begin{align}
	P_p&=\frac{1}{2{\rm Re}A_p} \nonumber \\
	&\propto\frac{1+p^2\tau^2}{p^3} \ , \label{eq:power_spectrum}
\end{align}
and the 1-loop diagram of decoherence rate $\int_{{\bf k}+{\bf k}'=-{\bf q}}P_kP_{k'}$ has UV divergence from the $\frac{\tau^2}{k}$ part of the power spectrum, suggesting us to renormalize the coefficient of the wave functional to resolve the UV problem. 

From (\ref{eq:on-shell-action}), the coefficient at the UV environment modes at $k\to+\infty$ is $A_k\propto k$, and we may construct a boundary counterterm to eliminate this while preserving the correct scaling for the system's super-horizon power spectrum $A_q\propto q^3$ when $q\to 0$, suggesting that \footnote{One may consider the counterterms without time derivative on the boundary like $\zeta^2$, $(\partial_i\zeta)^2$ and $(\partial^2\zeta)^2$, but these only change the quadratic phase of the wave functional, and thus it cannot resolve the divergence in decoherence. Similarly, in \cite{Burgess:2022nwu}, the UV divergence appears in the imaginary part of the Lindblad equation which is not related to decoherence.}
\begin{align}
	\delta S_{2}(\zeta)&=\int_{\partial M}d^3y \ 2\epsilon M_p^2 a^3\frac{1}{a^2H^2} \partial_i\dot{\zeta}\partial_i\zeta +\mathcal{O} \left(\frac{\partial^4}{a^4H^4}\right) \ , \label{eq:gauss_counter}
\end{align}
where the $\mathcal{O} \left(\partial^4\right)$ term will also be considered later, and the coefficient is changed to
\begin{align}
	A_p+\delta A_p \propto \frac{-ip^2+p^3\tau-2ip^4\tau^2+2p^5\tau^3}{\tau(1+p^2\tau^2)} \ .
\end{align}
The coefficient $A_p\propto p^3$ when $p\to +\infty$, so the $\frac{1}{p}$ part in the power spectrum (\ref{eq:power_spectrum}) is eliminated, whereas the super-horizon power spectrum $\propto \frac{1}{p^3}$ is unaffected. As discussed in \cite{Burrage:2011hd}, the boundary counterterm (\ref{eq:gauss_counter}) involving $\dot{\zeta}$ corresponds to a field redefinition 
\begin{align}
	\zeta_{\bf p}\to\bar{\zeta}_{\bf p}=\left(1+\frac{p^2}{a^2 H^2}\right)\zeta_{\bf p} \ , \label{eq:field_redefinition}
\end{align}
and the corresponding counterterms including the $\mathcal{O}(\partial^4)$ contribution are given by the action difference after redefining the field $\delta S_2=S_2\left(\bar{\zeta}\right)-S_2\left(\zeta\right)$. It is clear that the redefinition dominates for sub-horizon modes, whereas the effect on super-horizon modes is suppressed. On the other hand, the cubic boundary term $\partial_t(a^3\zeta)$ introduces additional terms through the field redefinition 
\begin{align}
	\partial_t\left(a\zeta \partial_i\zeta \partial_i \zeta\right) \ , \ \partial_t\left[\frac{1}{a}\partial^2\zeta(\partial_i \zeta)^2\right] \ , \ \partial_t\left(\frac{1}{a^3}\partial^2\zeta\partial^2\zeta\partial^2\zeta\right) \ ,
\end{align}
where the first two types already exist on the boundary \cite{Burrage:2011hd}, and they are suppressed by the scale factor $a(t)$ compared to the term $\partial_t(a^3\zeta^3)$, whereas the last term proportional to the system-mode physical momentum $\frac{q^2}{a^2}$ is further suppressed, so we will not consider these terms in the decoherence problem.

After the field redefinition, the on-shell action becomes
\begin{align}
	S_2+\delta S_{2}&=\int \frac{d^3p}{(2\pi)^3} \ \epsilon M_p^2 a^3 \frac{\dot{\bar\zeta}_{\bf p}}{\bar\zeta_{\bf p}} \left(1+\frac{p^2}{a^2 H^2}\right)^2\zeta_{\bf p}\zeta_{-{\bf p}}+\int d^4x \ \frac{\bar{\zeta}}{2}  \frac{\delta L_2}{\delta \bar{\zeta}}(\bar{\zeta}) \nonumber \\
	&=\frac{i}{2}\int \frac{d^3p}{(2\pi)^3}\left(1+\frac{p^2}{a^2 H^2}\right)^2 A_p  \zeta_{\bf p}\zeta_{-{\bf p}} \ ,
\end{align}
where we use the fact that the new field $\bar{\zeta}$ satisfies the EOM in the last line. In the literature, field redefinitions depending on comoving momenta have been applied to renormalize the wave functional on boundary \cite{Cespedes:2020xqq}, and we will show that the similar redefinition can resolve the UV divergence in the decoherence rate. The field redefinition (\ref{eq:field_redefinition}) changes the coefficient $A_p$ and power spectrum $P_q$ to
\begin{align}
	A_p &= \frac{2\epsilon M_p^2}{H^2}\frac{-ip^2+p^3\tau-ip^4\tau^2+p^5\tau^3}{\tau} \nonumber \\
	P_p &=\frac{H^2}{4\epsilon M_p^2} \frac{1}{p^3(1+p^2\tau^2)} \ ,
\end{align}
so the UV limit of the power spectrum is suppressed as $P_p\propto \frac{1}{p^5}$, causing no UV divergence in the 1-loop decoherence rate 
\begin{align}
	&\Gamma_\zeta^{\rm bd}(q,\tau) \nonumber \\
	&=\frac{4\pi^2\Delta^2_\zeta}{q^3}|\mathcal{F}|^2 \frac{1}{4\pi^2}\left(\int^{+\infty}_{k_{\rm min}, \ |k-q|>k_{\rm min}} dk \int^1_{-1}dx-\int^{2q+k_{\rm min}}_{k_{\rm min}, \ |k-q|>k_{\rm min}}dk \int^{\frac{k^2+q^2-(q-k_{\rm min})^2}{2kq}}_{\frac{k^2+q^2-(q+k_{\rm min})^2}{2kq}} dx\right) \nonumber \\
	&\times k^2P_\zeta(k)P_\zeta(\sqrt{k^2+q^2-2kqx})\nonumber \\
	&= \frac{729 \Delta_{\zeta} ^2}{16  \epsilon ^2} \left(\frac{aH}{q}\right)^6  J\left(\frac{q}{aH},\Delta N\right) \nonumber \\
	&\approx  \frac{729 \Delta_{\zeta} ^2}{16  \epsilon ^2}\Bigg[4 \left(\Delta N-1\right) \left(\frac{aH}{q}\right)^6-4 \Delta N \left(\frac{aH}{q}\right)^4+\frac{5 \pi }{2} \left(\frac{aH}{q}\right)^3 
	+\left(4 \Delta N-7\right) \left(\frac{aH}{q}\right)^2 \Bigg] \nonumber \\ &+\mathcal{O}\left(\frac{aH}{q}\right) \ , \label{eq:renormalized_deco_boundary}
\end{align}
where the complicated function $J\left(Q,\Delta N \right)$ is shown in the appendix (\ref{eq:function_J}), and we expand it with the late-time limit in the last line.

\section{Decoherence of inflaton in the semi-classical gravity}
\label{sec:inflaton_self-interaction}
In the semi-classical gravity, the metric cannot include any quantum fluctuation, and the conservation of $\zeta$ on super-horizon scales does not imply the frozen quantum degree of freedom. For such a semi-classical theory, all the geometric quantities, including coordinates and temporal boundaries, remain classical, and we thus select the inflaton fluctuation $\varphi$ to describe the quantum fluctuation. The main difference between the $\delta\phi$-gauge of the non-classical gravity and such a choose in the semi-classical gravity is the evaluation of the ADM constraints: the former are quantum operator equations, whereas the latter are semi-classical equations \footnote{As discussed in \cite{Maldacena:2002vr}, the first-order $N$ and $N^i$ are sufficient for the cubic interaction.}
\begin{align}
N-1&=\left\langle\frac{\dot{\phi}}{2H}\varphi\right\rangle \nonumber \\
&=0 \ , \nonumber \\ 
 N^i&=\left\langle\frac{{\dot{\phi}}^2}{2H^2}\partial_i\partial^{-2}\frac{d}{dt}\left(-\frac{H}{\dot{\phi}}\varphi\right) \right\rangle \nonumber \\
 &=0 \ .
\end{align}
This implies that the inflaton is a spectator field in this case, and the leading cubic interaction is the inflaton's self-interaction.

The leading gravity-independent self-interaction is the inflaton $V'''$ term
\begin{align}
\mathcal{L}_{\rm int}=-\frac{a^3}{6}V'''\varphi^3 \ , \label{eq:V3primes}
\end{align} 
and the size of $V'''$ can be estimated as
\begin{align}
V'''=\frac{H^3}{\dot{\phi}}\epsilon  \left(-12 \epsilon +9 \eta +4 \epsilon ^2-9 \epsilon  \eta +3 \eta ^2\right) \ , \label{eq:Vtripleprime}
\end{align}
with the interaction Hamiltonian in the comoving momentum space is
\begin{align}
\tilde{H}^{\rm (int)}_{\bf{k},\bf{k}',\bf{q}}=\frac{a^3}{6}V''' . \label{eq:H_int_inflaton}
\end{align} 
With this interaction, we can check the decoherence caused by the inflaton, independent to the non-classicality of gravity.

\subsection{Decoherence from sub-horizon environment}
For simplicity, we assume that the inflaton's effective mass can be neglected and the decoherence caused by the sub-horizon environment can be approximated by the squeezed limit $q\ll k\approx k'$. With (\ref{eq:H_int_inflaton}), the exponent of the non-Gaussian part of wave functional (\ref{eq:F_definition}) is
\begin{align}
\mathcal{F}_{\bf{k},\bf{k}',\bf{q}}\approx-\frac{V'''}{6 H^4 (1-ik \tau)^2 }I_k(\tau) \ , \label{eq:F}
\end{align}
where 
\begin{align}
I_k(\tau)=-\frac{1}{3} i \left\{2 k^3 e^{-2 i k \tau } [\pi -i \text{Ei}(2 i k \tau )]+\frac{1+k \tau  (k \tau -2 i)}{\tau ^3}\right\}\ .
\end{align}
The decoherence rate (\ref{eq:Gamma_deco}) from the sub-horizon environment is thus
\begin{align}
\Gamma^{\rm sub}_{\varphi} &=-\frac{4\pi^2 \Delta_{\varphi} ^2 {V'''}^2}{  H^4 q^3 \tau ^3} C\nonumber \\
&=4\pi^2 C \epsilon^2  \left(-12 \epsilon +9 \eta +4 \epsilon ^2-9 \epsilon  \eta +3 \eta ^2\right)^2\Delta_{\zeta} ^2 \left(\frac{aH}{q}\right)^3 , \label{eq:deco_sub_self}
\end{align}
where $C$ is the integral of tracing out the modes with $-k\tau>1$
\begin{align}
C&=\int^{-1}_{-\infty}dz\ \frac{z^2 \left(z^2+1\right)^2 \left| \frac{\frac{z (z-2 i)+1}{z^3}+2 e^{-2 i z} [\pi -i \text{Ei}(2 i z)]}{(1-i z)^2}\right| ^2}{2592 \pi
   ^2} \nonumber \\
   &\approx 1.2\times 10^{-4}\ .
\end{align}
In the last line of (\ref{eq:deco_sub_self}), we set the size of $\dot{\phi}$ in $V'''$ (\ref{eq:Vtripleprime}) with the relation
\begin{align}
\dot{\phi}^2&=H^2\frac{\Delta^2_{\varphi}}{\Delta^2_{\zeta}} \ ,
\end{align}
for letting the inflaton's $V'''$ term has the same strength as the one in the non-classical gravity, and it should not be understood as $\varphi=-\frac{\dot{\phi}}{H}\zeta$ in the semi-classical gravity.

\subsection{Decoherence from the super-horizon environment}
For super-horizon system and environment modes, ${\rm Max}\{ q,k,k'\}\ll-\tau^{-1}$, we expand the integrand up to $\mathcal{O}\left(\tau^{-2}\right)$
\begin{align}
&P_kP_{k'}|\mathcal{F}_{{\bf k},{\bf k}',{\bf q}}|^2 \nonumber \\
&\approx \frac{V'''^2}{36H^4}\left[\frac{1}{36 k^3 k'^3 \tau ^6}+\frac{3 k^2+3 k'^2+2 q^2}{36 k^3  k'^3\tau ^4}+\frac{\pi 
   \left(k^3+k'^3+q^3\right)}{36 k^3  k'^3\tau ^3}+\frac{3 k^4+3 k^2 k'^2+3 k'^4+q^4}{36
   k^3 k'^3\tau ^2 }\right]\nonumber \\
   &+\mathcal{O}(\tau^{-1}) \ ,
\end{align}
as the most sub-dominated term by the $\zeta$ bulk interaction (\ref{eq:L_int_cubic_usual}) is proportional to $a(t)^2$.
Similar to the decoherence of $\zeta$, the case of inflaton also receives IR divergence from the super-horizon modes, and we apply the method in Sect.~\ref{sec:divergences} to resolve this. The super-horizon decoherence rate is
\begin{align}
\Gamma^{\rm super}_{\varphi}&\approx\frac{\Delta_\zeta ^2 \epsilon ^2 \left[3 \eta  (\eta +3)+4 \epsilon ^2-3 (3 \eta +4) \epsilon \right]^2}{3888 } \Bigg\{12\left( \Delta N-1\right)\left(\frac{aH}{q}\right)^6+12\left(5 \Delta N-2\right) \left(\frac{aH}{q}\right)^4\nonumber \\
&+4\left[ -6 \pi  \Delta N-3 \pi  \log \left(\frac{aH}{q}\right)+6 \pi +4\right]\left(\frac{aH}{q}\right)^3+3\left( 16 \Delta N-25\right)\left(\frac{aH}{q}\right)^2\Bigg\} \nonumber \\
&+\mathcal{O}\left(\frac{k_{\rm min}}{q},\frac{aH}{q}\right) \ , \label{eq:deco_phi_super}
\end{align}
and the total decoherence rate of inflaton is $\Gamma_{\varphi}=\Gamma^{\rm sub}_\varphi+\Gamma^{\rm super}_\varphi$.

\subsection{Compare with the decoherence rate of $\zeta$}
\begin{figure}
\centering
\includegraphics[width=0.9\textwidth]{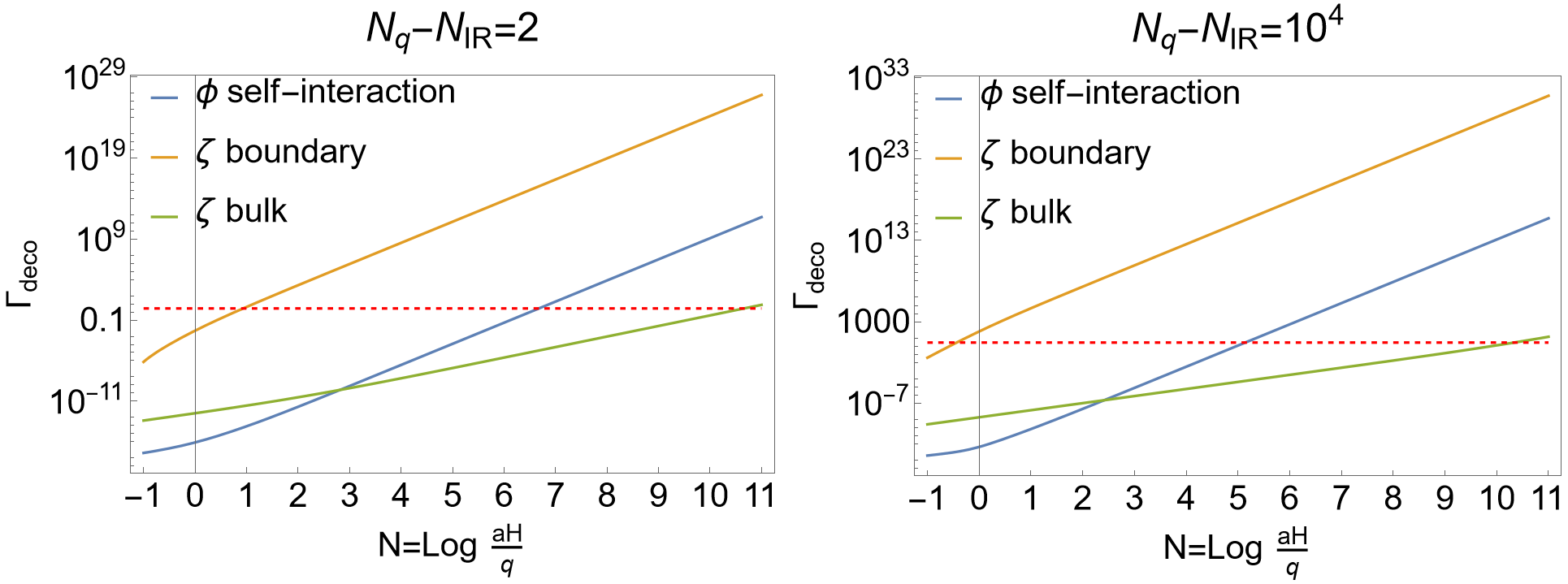}
\caption{The comparison between $\Gamma_{\varphi}$, $\Gamma^{\rm bd}_\zeta$ and $\Gamma^{\rm bulk}_\zeta$ with the IR cutoff set to be the minimum $\Delta N\approx 2$ and $\Delta N\approx 10^4$ used in \cite{Martin:2018zbe} respectively. The red-dashed line indicates the value $\Gamma=1$ when the field starts to decohere, and $N=\log\left(\frac{aH}{q}\right)$ is the e-folds elapsed after crossing the horizon. Since $\Gamma^{\rm bd}_\zeta=1$ around the horizon crossing, the full form of $J\left(Q,\Delta N\right)$ (\ref{eq:function_J}) without making the late-time expansion (\ref{eq:renormalized_deco_boundary}) is used.}
\label{fig:compare_deco}
\end{figure}

We are ready to compare the decoherence rate of $\varphi$ with $\zeta$ from either the boundary term (\ref{eq:renormalized_deco_boundary}) or the bulk Lagrangian (\ref{eq:L_int_cubic_usual}) which has been calculated in \cite{Nelson:2016kjm} \footnote{The factor $\frac{4\pi^2}{8\pi}$ comes from the change of coefficient in (\ref{eq:Gamma_deco}).}
\begin{align}
\Gamma_\zeta^{\rm bulk}=\frac{4\pi^2}{8\pi}\left[\left(\frac{\epsilon+\eta}{12}\right)^2\Delta^2_\zeta\left(\frac{aH}{q}\right)^3+\frac{(\epsilon+\eta)^2\Delta^2_\zeta}{9\pi}\left(\frac{aH}{q}\right)^2\left(\Delta N-\frac{19}{48}\right)\right] \ ,
\end{align}
where the first term is the dominated contribution from the sub-horizon environment, whereas the sub-dominated second term is attributed to the super-horizon environment. Using the observed values \cite{Planck:2018jri}
\begin{align}
\Delta^2_{\zeta}\approx 2.5\times 10^{-9}\ , \ \epsilon<0.006 \ , \ \eta\approx 0.03 \ ,
\end{align}
Fig. \ref{fig:compare_deco} shows the decoherence rates of inflaton $\varphi$ by the self-interaction and $\zeta$ by the bulk and boundary terms, and $\Gamma_\zeta^{\rm bd}$ dominates before and after the decoherence happens ($\Gamma\sim 1$). Comparing with (\ref{eq:renormalized_deco_boundary}) and (\ref{eq:deco_phi_super}), the super-horizon contribution with the bulk interaction of $\zeta$ (\ref{eq:L_int_cubic_usual}) is suppressed by $a(t)^{-4}$ since each spatial Laplacian introduces $a(t)^{-2}$, and this makes the inflaton $\varphi$ decoheres earlier than $\zeta$ by the bulk interaction although the self-interaction of inflaton is more slow-roll suppressed, showing the importance of considering the super-horizon environment.

\section{Conclusion and outlook}
\label{sec:conclusion}
We have studied the boundary term of the comoving curvature perturbation $\zeta$, emphasizing its importance in the variational principle and the decoherence of super-horizon modes. This boundary term cannot contribute to the equation of motion of $\zeta$ and correlators of the form $\langle\zeta^n\rangle$ as it is a total derivative, so it is neglected in literature. However, such a term is necessary to exist for making the variation well-defined, as shown in the ADM formalism and the Einstein-Hilbert action with the Gibbons-Hawking-York term, and it is unique, provided that the fixed quantity on the boundary is the induced metric $h^{ij}$. To show the effect caused by the term, we study the decoherence of cosmological perturbations by partitioning them into a system and an environment according to their comoving momenta. We calculated the decoherence rate of the observable super-horizon modes with improving the method of Schr\"odinger wave functional in \cite{Nelson:2016kjm} to include the non-Gaussian phase from the boundary term. Technically, the decoherence rate were evaluated by the perturbation method and the saddle-point approximation, and the preliminary 1-loop result was shown to have IR and UV divergences. The IR divergence is resolved by applying the duration of inflation as a real cutoff, and the UV divergence is resolved by the renormalizing the wave functional with field redefinition.

The boundary term of $\zeta$ was shown to produce much larger decoherence effect than the leading bulk term of $\zeta$ and the self-interaction of inflaton $\varphi$, providing us a possibility to distinguish the non-classical and semi-classical gravity through the decoherence. Our results show that $\zeta$ starts to decohere around the horizon crossing, which improves the previous estimation in \cite{Nelson:2016kjm}, whereas $\varphi$ and the one from the $\zeta$ bulk term need at least 5-6 and 10 e-folds respectively. For the non-classical gravity, $\zeta$ can describe the quantum fluctuations, and its decoherence from the minimal gravitational non-linearity already happens much faster than the leading decoherence effect in the semi-classical gravity, caused by the self-interaction of the quantum degree of freedom $\varphi$. Our results thus provide parameter regions to distinguish the non-classical and semi-classical gravity, such as probing the quantumness of the modes which have crossed the horizon for more than 1 e-fold but less than 6 e-folds.

In the discussion of the boundary term with the $\delta\phi$-gauge in Sec. \ref{sec:deltaphi_gauge_boundary}, we have pointed out that the choice of boundary is an important factor to determine the decoherence. For probing observables frozen outside the horizon, the constant-time hypersurface defined in the $\zeta$-gauge is a natural choice due to the conservation of super-horizon modes of $\zeta$, and the large decoherence effect caused by the boundary term supports the classicality of such observables. For observables favored by quantum-information aspects, small decoherence effect from the boundary term is desired, and we have shown that the constant-time hypersurface in $\delta\phi$-gauge can achieve this. Another reason for choosing $\varphi$ to be the fluctuation in the quantum-information tests is its non-vanishing conjugated momentum caused by the non-linear evolution, supporting the probes of non-commuting operators. We plan to address this observation in more details in a future work. 

\section*{Acknowledgments}
We thank Xi Tong for helpful discussion. This work was supported in part by the National Key R\&D Program of China (2021YFC2203100), the NSFC Excellent Young Scientist Scheme (Hong Kong and Macau) Grant No. 12022516, and by the RGC of Hong
Kong SAR, China (Project No. CRF C6017-20GF and 16306422).

\appendix
\section{Compare the in-in formalism with the Schr\"odinger wave functional}
This appendix shows the calculation of correlators with two methods: the in-in formalism and the Schr\"odinger wave functional, and here for convenience we do not distinguish the comoving momenta for system and environment, $\bf q$ and $\bf k$, which can be used to label all the modes of $\zeta$.

\label{sec:appendix_compare_inin_schrodinger}
\subsection{In-in correlators}
For curvature perturbation $\zeta$, it also includes non-derivative cubic term, but it is in a total time derivative
\begin{align}
H_{\rm bd}=M_p^2\int d^3x \ \partial_t\left(9a^3H\zeta^3\right) \ .
\end{align}
Its first-order contribution to correlators is obtained by the in-in formalism
\begin{align}
\langle\Omega|O(\tau)|\Omega\rangle_1 &=2{\rm Im}\left[\int^t_{t_0}dt_1\langle 0|O^I(\tau)H^I_{\rm bd}(t_1)|0\rangle\right] \nonumber \\
&=18a^3HM_p^2{\rm Im}\left[\int_{{\bf k},{\bf k}',{\bf q}}\langle 0|O^I(\tau)\zeta_{{\bf k}}(\tau)\zeta_{{\bf k}'}(\tau)\zeta_{{\bf q}}(\tau)|0\rangle\right] \ ,
\end{align}
where $\zeta$ is in the interaction picture. For $O(\tau)=\zeta_{{\bf k}_1}\zeta_{{\bf k}_2}\zeta_{{\bf k}_3}$, the result in the square bracket is proportional to $(2\pi)^3\delta^{(3)}({\bf k}_1+{\bf k}_2+{\bf k}_3)|u_{k_1}|^2|u_{k_2}|^2|u_{k_3}|^2$,
and therefore it does not contribute to the 3-point correlation of $\zeta$. However, the non-Gaussianity of the boundary term can be reflected by the correlation of conjugate momenta, and let $O(\tau)={\dot\zeta}_{{\bf k}_1}{\dot\zeta}_{{\bf k}_2}{\dot\zeta}_{{\bf k}_3}$
\begin{align}
\langle\Omega|{\dot\zeta}_{{\bf k}_1}{\dot\zeta}_{{\bf k}_2}{\dot\zeta}_{{\bf k}_3}|\Omega\rangle'_1 &=18H(3!)M_p^2{\rm Im}\left(\prod_{i=1}^3 u'_{k_i}u^*_{k_i}\right) \nonumber \\
&=\frac{27 H^7 \tau ^4 \left(k_1 k_2 k_3 \tau ^2-k_1-k_2-k_3\right)}{16 k_1 k_2 k_3 M_p^4 \epsilon ^3} \ , \label{eq:3con_mom_in_in}
\end{align}
where the factor of momentum conservation is not written. The correlation does not vanish identically and converges to zero at $\tau\to 0$.

\subsection{Wave functional approach}
The purely imaginary non-Gaussian exponent implies that the 3-point correlation of $\zeta$ vanishes at all order
\begin{align}
\langle \zeta_{{\bf k}_1}\zeta_{{\bf k}_2}\zeta_{{\bf k}_3}\rangle&=\int D\zeta  \  |\Psi(\zeta)|^2  \zeta_{{\bf k}_1}\zeta_{{\bf k}_2}\zeta_{{\bf k}_3} \nonumber \\
&=\int D\zeta  \  |\Psi_G(\zeta)|^2  \zeta_{{\bf k}_1}\zeta_{{\bf k}_2}\zeta_{{\bf k}_3}  \nonumber \\
&=0 \ ,
\end{align}
whereas the correlation of conjugate momenta is
\begin{align}
\langle {\dot\zeta}_{{\bf k}_1}{\dot\zeta}_{{\bf k}_2}{\dot\zeta}_{{\bf k}_3} \rangle &= \left(\frac{1}{2\epsilon M_p^2a^3}\right)^3\langle\Pi_{-{\bf k}_1}\Pi_{-{\bf k}_2}\Pi_{-{\bf k}_3} \rangle \nonumber \\
&=i \left(\frac{1}{2\epsilon M_p^2a^3}\right)^3 \int D\zeta \ \Psi^*(\zeta)\frac{\delta}{\delta \zeta_{-{\bf k}_1}}\frac{\delta}{\delta \zeta_{-{\bf k}_2}}\frac{\delta}{\delta \zeta_{-{\bf k}_3}}\Psi(\zeta) \ .
\end{align}
The triple functional derivative is
\begin{align}
&\left[\Psi(\zeta)\right]^{-1}\frac{\delta}{\delta \zeta_{-{\bf k}_1}}\frac{\delta}{\delta \zeta_{-{\bf k}_2}}\frac{\delta}{\delta \zeta_{-{\bf k}_3}}\Psi(\zeta) \nonumber\\ &=(2\pi)^3\delta^{(3)}\left( {\bf k}_1+{\bf k}_2+{\bf k}_3\right) 3! \mathcal{F}\nonumber \\
&+3!\sum_{i=1}^3\mathcal{F}\zeta_{-{\bf k}_i}\left(-A_{k_i}\zeta_{{\bf k}_i}+3\int_{{\bf k}+{\bf k}'={\bf k}_i}\mathcal{F}\zeta_{\bf k}\zeta_{{\bf k}'}\right) \nonumber \\
&+\prod_{i=1}^3\left(-A_{k_i}\zeta_{{\bf k}_i}+3\int_{{\bf k}+{\bf k}'={\bf k}_i}\mathcal{F}\zeta_{\bf k}\zeta_{{\bf k}'}\right) \nonumber \\
&=3!\mathcal{F}\Bigg[(2\pi)^3\delta^{(3)}\left({\bf k}_1+{\bf k}_2+{\bf k}_3\right)-\left(\sum_i A_{k_i}\zeta_{{\bf k}_i}\zeta_{-{\bf k}_i}\right) \nonumber \\
&+\left(\frac{1}{2}A_{k_1}A_{k_2}\zeta_{{\bf k}_1}\zeta_{{\bf k}_2}\int_{{\bf k}+{\bf k}'={\bf k}_3}\mathcal{F}\zeta_{\bf k}\zeta_{{\bf k}'}+2{\rm  \ perms}\right)\Bigg]+\mathcal{O}(|\mathcal{F}|^2,\zeta^{2n+1}) \ ,
\end{align}
where here the coefficient of the non-Gaussian exponent
\begin{align}
\mathcal{F}&=-9iM_p^2a^3H \ ,
\end{align}
and the terms of $\mathcal{O}(|{\mathcal{F}|^2})$ and odd orders of $\zeta$ are neglected in the last line. Using the relation
\begin{align}
\langle\zeta_{{\bf k}}\zeta_{{\bf k}'}\rangle&=(2\pi)^3\delta^{(3)}({\bf k}+{\bf k}')\frac{1}{2{\rm Re}A_k(\tau)} \nonumber \\
&=(2\pi)^3\delta^{(3)}({\bf k}+{\bf k}')P_k(\tau) \ ,
\end{align}
the correlation of conjugate momenta is obtained
\begin{align}
\langle {\dot\zeta}_{{\bf k}_1}{\dot\zeta}_{{\bf k}_2}{\dot\zeta}_{{\bf k}_3} \rangle'&=\frac{27 H^7 \tau ^4 \left(k_1 k_2 k_3 \tau ^2-k_1-k_2-k_3\right)}{16 k_1 k_2 k_3 \epsilon ^3 M_p^4} \ ,
\end{align}
which agrees with the in-in formalism (\ref{eq:3con_mom_in_in}).

\section{The function $J(Q,\Delta N)$ in the 1-loop decoherence rate}
\begin{figure}
	\centering
	\includegraphics[width=0.7\textwidth]{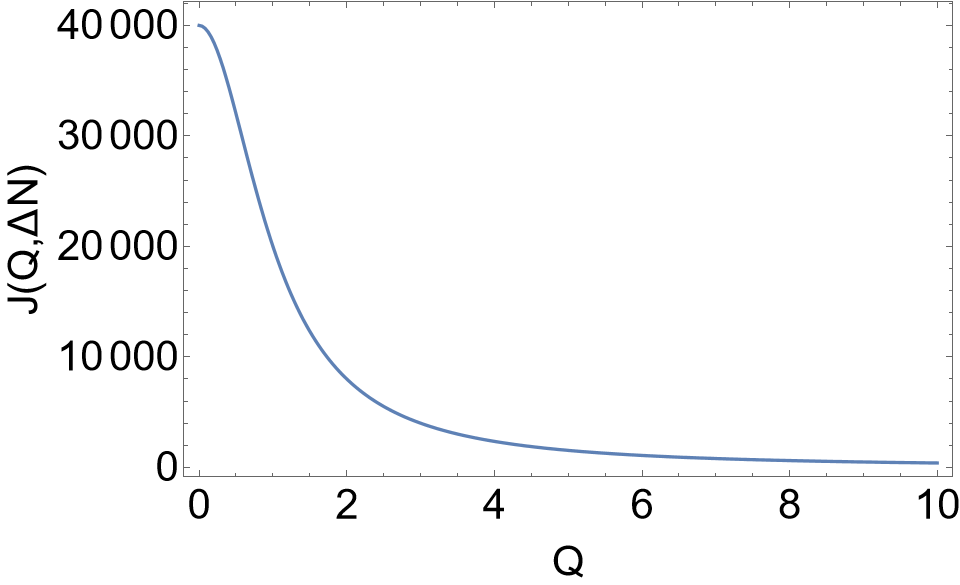}
	\caption{The function $J\left(Q,\Delta N\right)$ with $N_q-N_{\rm IR}=10^4$. \label{fig:function_J}}
\end{figure}

The complicated function $J\left(Q,\Delta N\right)$ is
\begin{align}
	&J\left(Q,\Delta N\right) \nonumber \\
	&=\frac{1}{8(1+Q^2)}\Bigg[-4 \pi  \tan ^{-1}(Q) Q^4-3 i \pi  \log (2 i-Q) Q^4+4 i \pi  \log (i Q+1) Q^4-4 i \pi  \log (i Q+2) Q^4 \nonumber \\
	&-4 \log (i Q+1) \log \left(\frac{i}{Q-2
		i}\right) Q^4+4 \log (i Q+2) \log \left(\frac{i}{Q-2 i}\right) Q^4 \nonumber \\
	&-2 \log (i-Q) \log \left(-\frac{i Q}{Q-2 i}\right) Q^4-2 \log (i Q+1)
	\log \left(-\frac{i Q}{Q-2 i}\right) Q^4+2 i \pi  \log \left(-\frac{i Q}{Q-2 i}\right) Q^4 \nonumber \\
	&-2 \log (i Q-1) \log \left(\frac{i Q}{Q+2
		i}\right) Q^4-2 \log (Q+i) \log \left(\frac{i Q}{Q+2 i}\right) Q^4+2 i \pi  \log \left(\frac{i Q}{Q+2 i}\right) Q^4 \nonumber \\
	&+3 i \pi  \log (Q+2 i)
	Q^4-2 \text{Li}_2\left(-\frac{1}{Q^2}\right) Q^4+4 \text{Li}_2\left(\frac{Q-i}{Q}\right) Q^4+4 \text{Li}_2\left(\frac{Q+i}{Q}\right) Q^4 \nonumber \\
	&+4
	\text{Li}_2\left(\frac{i}{Q+2 i}\right) Q^4-4 \text{Li}_2\left(\frac{Q+i}{Q+2 i}\right) Q^4-8 \text{Li}_2\left(1+\frac{i}{Q-2 i}\right)
	Q^4+\frac{2 \pi ^2 Q^4}{3}-16 \tan ^{-1}(Q) Q^3 \nonumber \\
	&+8 i \log (i-Q) Q^3-8 i \log (Q+i) Q^3+16 \pi  Q^3-4 \pi  \tan ^{-1}(Q) Q^2-3 i \pi  \log (2
	i-Q) Q^2 \nonumber \\
	&+4 i \pi  \log (i Q+1) Q^2-4 i \pi  \log (i Q+2) Q^2-8 \log (Q+i) Q^2-4 \log (i Q+1) \log \left(\frac{i}{Q-2 i}\right) Q^2 \nonumber \\
	&+4 \log
	(i Q+2) \log \left(\frac{i}{Q-2 i}\right) Q^2-2 \log (i-Q) \log \left(-\frac{i Q}{Q-2 i}\right) Q^2 \nonumber \\
	&-2 \log (i Q+1) \log \left(-\frac{i
		Q}{Q-2 i}\right) Q^2+2 i \pi  \log \left(-\frac{i Q}{Q-2 i}\right) Q^2-2 \log (i Q-1) \log \left(\frac{i Q}{Q+2 i}\right) Q^2 \nonumber \\
	&-2 \log (Q+i)
	\log \left(\frac{i Q}{Q+2 i}\right) Q^2+2 i \pi  \log \left(\frac{i Q}{Q+2 i}\right) Q^2+3 i \pi  \log (Q+2 i) Q^2 \nonumber \\
	&-8 \log
	\left((i-Q) \left(Q^2+1\right)\right) Q^2-2 \text{Li}_2\left(-\frac{1}{Q^2}\right) Q^2+4
	\text{Li}_2\left(\frac{Q-i}{Q}\right) Q^2+4 \text{Li}_2\left(\frac{Q+i}{Q}\right) Q^2 \nonumber \\
	&+4 \text{Li}_2\left(\frac{i}{Q+2 i}\right) Q^2-4
	\text{Li}_2\left(\frac{Q+i}{Q+2 i}\right) Q^2-8 \text{Li}_2\left(1+\frac{i}{Q-2 i}\right) Q^2+\frac{2 \pi ^2 Q^2}{3}+8 i \pi  Q^2-32 Q^2 \nonumber \\
	&+32(\Delta N-1) \Bigg]\ . \label{eq:function_J}
\end{align}
Although it looks like a complex function, it has real and positive values in the region related to our problem, as shown in Fig. \ref{fig:function_J}. Since the decoherence caused by the boundary term happens around the horizon crossing $Q=\frac{q}{aH}=1$, the full form of $J\left(Q,\Delta N\right)$ without making the late-time expansion should be kept when we compare the decoherence rates.

\bibliographystyle{utphys}
\bibliography{reference}

\end{document}